\documentclass[twocolumn,aps,showpacs,prb,tightenlines,amsmath,amssymb,superscriptaddress]{revtex4}
\usepackage{graphicx}
\usepackage{amssymb}
\usepackage{dcolumn}
\usepackage{amsmath}
\usepackage{bm}
\usepackage{colordvi}
\usepackage{mathrsfs}
\makeatletter

\newcommand{\Rmnum}[1]{\expandafter\@slowromancap\romannumeral #1@}
\makeatother

\begin{document}

\title{Hole spin relaxation in bilayer WSe$_{\bm 2}$}
\author{F. Yang}
\affiliation{Hefei National Laboratory for Physical Sciences at
Microscale and Department of Physics,
University of Science and Technology of China, Hefei,
Anhui, 230026, China}

\author{L. Wang}
\affiliation{Hefei National Laboratory for Physical Sciences at
Microscale and Department of Physics,
University of Science and Technology of China, Hefei,
Anhui, 230026, China}

\author{M. W. Wu}
\thanks{Author to whom correspondence should be addressed}
\email{mwwu@ustc.edu.cn.}
\affiliation{Hefei National Laboratory for Physical Sciences at
Microscale and Department of Physics, University of Science and
Technology of China, Hefei, Anhui, 230026, China}
\affiliation{Department of Physics, Kyushu University, 
6-10-1 Hakozaki, Fukuoka, 812-8581, Japan}

\date{\today}

\begin{abstract} 
We investigate the hole spin relaxation due to the Rashba spin-orbit coupling
induced by an external perpendicular electric field in bilayer WSe$_2$. The
Rashba spin-orbit coupling coefficients in bilayer WSe$_2$ are constructed from
the corresponding monolayer ones. In contrast to monolayer WSe$_2$, the
out-of-plane component of the bilayer Rashba spin-orbit coupling acts as a
Zeeman-like field with opposite directions but identical values in the two
valleys.  For in-plane spins, this Zeeman-like 
field, together with the intervalley hole-phonon scattering, opens an intervalley
spin relaxation channel, which is found to dominate the in-plane spin relaxation
in bilayer WSe$_2$ even at low temperature.  For out-of-plane spins, this
Zeeman-like field is superimposed by the identical Hartree-Fock effective
magnetic fields in the two valleys, and hence different total effective magnetic
fields between two valleys are obtained.  Owing to the large difference of the
total fields at large spin polarization,  different out-of-plane spin relaxation
times in the two valleys are obtained when the intervalley hole-phonon
scattering is weak at low temperature and low hole density. This difference
in the spin relaxation times can be suppressed by enhancing the intervalley
hole-phonon scattering through increasing temperature or hole density.
Moreover,  
at large spin polarization and low temperature, due to the weak intravalley
hole-phonon scattering but relatively strong hole-hole Coulomb scattering,  the
fast spin precessions are found to result in a quasi hot-hole Fermi distribution
characterized by an effective hot-hole temperature larger than the temperature,
which also enhances the intervalley scattering. During this
process, it is interesting to discover that the initially equal hole densities
in the two valleys are broken in the temporal evolution, and a valley
polarization is built up. It is further revealed that this comes from the
different spin relaxation processes at large spin polarization and different
spin-conserving intervalley scattering rates between spin-up and -down holes
due to the different effective hot-hole temperatures.

\end{abstract}

\maketitle 

\section{Introduction}
 
Very recently, the spin dynamics in monolayer (ML) transition metal 
dichalcogenides (TMDs) has attracted much
attention,\cite{s_1,s_2,s_3,s_4,s_5,s_6,s_7} partly due to their unique
electric\cite{ML_g1,ML_g2,ML_g3,ML_g4,ML_g5,ML_g6,ML_g7,ML_g8,splitting_1,splitting_2,splitting_3}
and novel optical\cite{ML_g7,ML_g8,msr_1,msr_2,msr_3,msr_4} properties. 
Specifically, owing to the direct gap at the K (K$'$)
point\cite{ML_g1,ML_g2,ML_g3,ML_g4} and large energy 
spin splitting of the valence bands,\cite{splitting_1,splitting_2,splitting_3}
the chiral optical valley selection rule in
ML TMDs \cite{msr_1,msr_2,msr_3,msr_4} allows optical control
of the valley pseudospin\cite{vp_1,vp_2,vp_3,vp_4,vp_5} and real
spin,\cite{ML_g4,ML_g5,ML_g6,ML_g7,splitting_1,splitting_2,msr_1,msr_2,msr_3,msr_4}
making them promising candidates for the spintronic application.

Among these studies in ML TMDs, carrier spin relaxation due to the
Elliot-Yafet\cite{Yafet,Elliott} 
(EY) and D'yakonov-Perel'\cite{DP} (DP) mechanisms
is an important property
to understand for any possible 
spintronic applications. For the intrinsic 
EY mechanism, it was claimed that the intervalley
out-of-plane spin-flip scattering is forbidden by
the time reversal symmetry, while the flexural phonon vibrations thermally 
activated at high temperature can lead to the intravalley out-of-plane spin
relaxation.\cite{s_1} Additionally, due to the marginal in-plane spin mixing,
the contribution of the EY mechanism to the in-plane spin
relaxation is negligible.\cite{s_5}
For the DP mechanism, in the intrinsic situation, the out-of-plane spin
relaxation processes are forbidden due the good quantum number of $\sigma_z$ but
the in-plane processes exist,\cite{s_4,s_5} including the intravalley and
intervalley processes. Particularly, the intervalley process arises from the
intrinsic spin-orbit coupling (SOC), which acts as 
opposite effective magnetic fields in the two valleys and hence opens an
intervalley in-plane spin relaxation channel in the presence of the intervalley
carrier-phonon scatterings. The extrinsic influence
such as the flexural
deformations,\cite{s_3} the external in-plane magnetic field,\cite{s_5} or the
Rashba SOC induced by an 
external perpendicular electric field,\cite{s_2} can cause the out-of-plane spin
relaxation.  Moreover, in contrast to the electron spin
relaxation, the hole spin relaxation processes are markedly suppressed due to
the large energy splitting of the valence bands.\cite{s_1,s_2,s_3}

Consisting of two-layer TMDs, bilayer (BL) TMDs also obey the chiral optical 
valley selection rule and possess good spin
characters,\cite{bandgap,electric1,electric2,tight-binding,Hamiltonian1,Hamiltonian2,inter-layer,Cui,exciton,Yu,Thick,pseudospin,optical,spin-orbit,electronic-structure,EM}  
apart from new features such as layer pseudospin,\cite{pseudospin} electrical-tuned magnetic
moments\cite{electric1} and magnetoelectric effect.\cite{electric2}    
Specifically, due to the $180^{\circ}$ in-plane rotation between the upper and
lower layers in BL TMDs, in the K (K$'$) valley, the hole bands in one
layer have opposite spin 
polarizations between the energy-degenerate ones in another layer. 
The interlayer hopping, which exists only in holes in the same valley
with the same spin, has no influence on this spin degeneracy. This is very
different from the ML TMDs, where the hole bands possessing opposite spin
polarizations in the K (K$'$) valley are largely energy split,
markedly suppressing the hole spin relaxation. In BL TMDs, with the
intrinsic EY
hole spin relaxation and the intrinsic DP in-plane one suppressed in each layer
while the intrinsic DP out-of-plane one forbidden, 
the Rashba SOC of the lowest two degenerate hole
bands\cite{electric2}   
\begin{equation}
\label{Rashba}
{\bf \Omega}^{\mu}=\big[-{\nu}(1+{\alpha}k^2)k_{y},{\nu}(1+{\alpha}k^2)k_{x},\mu\eta\big]E_{z},
\end{equation}
induced by an external perpendicular electric field $E_z$, is expected to make
the dominant contribution to the hole spin relaxation. 
Here, the ${\hat z}$-axis is set to be perpendicular to the BL TMD plane; $\mu=1
(-1)$ representing the K (K$'$) valley.   
The out-of-plane component of Rashba SOC serves as a
Zeeman-like term ${\mu\eta}E_{z}$ with opposite effective magnetic fields in the
two valleys.\cite{electric2,Hamiltonian1} Due to this coupling of real spin and valley
pseudospin, one may expect the interplay of the spin polarization with the
valley polarization and the rich physics of spin and valley dynamics
in BL TMDs.
Specifically, for in-plane spins, this Zeeman-like term, together
with the intervalley hole-phonon scattering,  opens a new intervalley spin
relaxation channel. For out-of-plane spins, this Zeeman-like term is
superimposed by the Coulomb Hartree-Fock (HF) self-energy, which has been
understood first theoretically\cite{magnetic_2,HF_1,HF_2} and then
experimentally\cite{HF_3} in semiconductors to serve as an effective magnetic
field 
\begin{equation}
\label{HF}
{\bf \Omega}_{\rm HF}({\bf k})={-\sum_{\bf k'}V_{{\bf k}-{\bf k'}}{\rm Tr}\big[\rho_{\bf k'}{\bm \sigma}\big]},
\end{equation}
with $V_{{\bf k}-{\bf k'}}$ being the screened Coulomb potential. In the case of
the valley-independent out-of-plane spin polarization, the total effective
magnetic fields 
\begin{equation}
\label{total}
{\bm \Omega}^{\mu}_{\rm eff}=({\Omega}_{\rm HF}+\mu{\eta}E_{z}){\bf e}_{\hat z} 
\end{equation}
have different values in the two valleys, leading to the valley-dependence of
the out-of-plane spin relaxation.
       
In the present work, by utilizing the kinetic spin Bloch equation (KSBE)
approach\cite{HF_2} with the hole-hole Coulomb,  (both the intra- and
intervalley) hole-phonon, and long-range hole-impurity scatterings included, we
investigate the hole spin relaxation in BL WSe$_2$ due to the 
Rashba SOC [Eq.~(\ref{Rashba})].  In our investigation, the initial
occupations of holes of each spin are identical in the two
valleys. In the case of small spin 
polarization (i.e., weak HF effective magnetic field), the  
total effective magnetic fields [Eq.~(\ref{total})] determined by the
Zeeman-like fields, have identical absolute values in the two valleys. We
investigate the temperature and hole density dependence of both 
the out-of- and in-plane spin relaxation times (SRTs), and show that for
in-plane spins, the relaxation process is dominated by the intervalley spin
relaxation channel induced by the Zeeman-like fields, and the
contribution of the intervalley hole-phonon scattering becomes dominant. But for
out-of-plane spin relaxation, the 
intervalley hole-phonon scattering is marginal, which indicates that
out-of-plane spins relax independently in the two valleys. With 
the intervalley scattering removed, the
out-of-plane SRTs $\tau_{sz}^{\mu}$ in the strong scattering
limit\cite{magnetic_1,magnetic_2}         
\begin{equation}
\label{tau_s}
{\tau_{sz}^{\mu}}=\frac{1+(\Omega^{\mu}_{\rm
    eff}\tau_p)^2}{\langle\Omega^2_{\perp}({\bf
    k})\rangle\tau_{p}}=\frac{1}{\langle\Omega_{\perp}^2({\bf
    k})\rangle\tau_{p}}+\frac{|\Omega^{\mu}_{\rm
    eff}|^2}{\langle\Omega^2_{\perp}({\bf k})\rangle}\tau_p,
\end{equation}
are identical in the two valleys due to the same absolute values of the total
effective 
magnetic fields. Here, ${\bm
  \Omega}_{\perp}({\bf k})$ and $\tau_p$ are the in-plane Rashba terms and
momentum relaxation time due to the intravalley scattering, respectively.

\begin{figure}[htb]
  {\includegraphics[width=8.5cm]{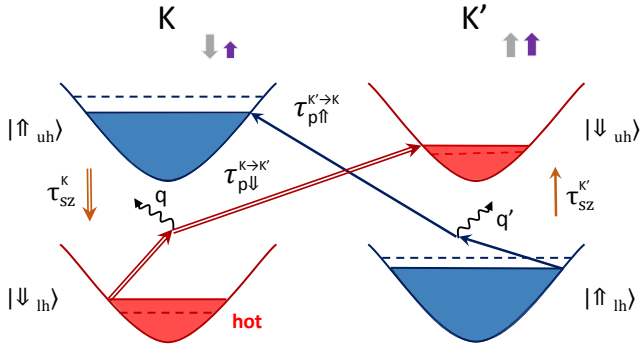}}
\caption{(Color online) Schematic of the different spin relaxation processes in
  the two valleys and valley polarization process. In the figure,
  $|\Uparrow(\Downarrow)~_{\rm uh (lh)}\rangle$ stands for the upper- (lower-)
  layer states for spin-up (-down) holes; the purple (gray) filled arrows, which have  
  the same (opposite) directions in the two valleys, denote the
  HF (Zeeman-like) effective magnetic fields; the dashed lines stand for the
  initial occupations of holes of each spin.  On one hand, this
  schematic shows that due to the smaller total effective magnetic field in the K
  valley, the SRT $\tau^{\rm K}_{sz}$ is faster than $\tau^{\rm K'}_{sz}$, which makes
  the density for spin-down (-up) holes in the K valley larger (smaller) than
  that in the K$'$ valley in the temporal evolution.  On other hand, the
  intervalley scattering time $\tau^{{\rm K}\rightarrow{\rm K'}}_{p~\Downarrow}$
  of spin-down holes is faster than that $\tau^{{\rm K'}\rightarrow{\rm
      K}}_{p~\Uparrow}$ of spin-up holes due to the larger effective hot-hole
  temperature for spin-down holes (red).   }   
\label{figyw1}
\end{figure}

In the case of large spin polarization (i.e., strong HF effective magnetic
field), for in-plane spins, the SRT is insensitive to the
    spin polarization, similar to the case of ML MoS$_2$.\cite{s_4} For out-of-plane
    spins, a large difference of the total effective magnetic fields
[Eq.~(\ref{total})] between the two valleys is obtained. We find that the
intervalley hole-phonon scattering in this situation is weak at low
temperature and low hole density, and hence the out-of-plane spins relax
independently in the two valleys, leading to different SRTs [Eq.~(\ref{tau_s})].
The enhancement of the intervalley hole-phonon 
scattering by increasing temperature can suppress this
difference in the SRTs. Moreover, the intervalley scattering also becomes
stronger with the increase of hole density at low temperature, which
originates from the different SRTs. Specifically, as
shown in Fig.~\ref{figyw1}, when the HF effective magnetic field and Zeeman-like
field have opposite directions, say, in the K valley, the smaller total
effective magnetic field [Eq.~(\ref{total})] leads to a 
faster SRT [Eq.~(\ref{tau_s})] in this valley. Over time, this faster spin
relaxation makes the density for spin-down (-up) holes larger (smaller)
than the corresponding one with the same spin in the K$'$ valley, triggering the
intervalley scattering of spin-down (-up) 
holes from the K (K$'$) valley to the K$'$ (K) one by emitting phonons to
suppress this density difference. With larger density
difference by increasing the hole density, the intervalley scattering
becomes stronger. In addition,  due to the weak intravalley hole-phonon
scattering at low temperature but relatively strong hole-hole Coulomb
scattering, we find that the fast spin precessions result in a quasi hot-hole
Fermi distribution characterized by an effective hot-hole temperature $T_{\rm
  eff}$ larger than the temperature $T$, which also enhances the intervalley
scattering.   

During above process, it is interesting to discover that the
initially equal hole densities in the two valleys are broken in the temporal
evolution, with more holes are accumulated in the K$'$ valley, leading to the
build up of the valley polarization. This arises from the larger
effective hot-hole 
temperature for spin-down holes than that for spin-up ones, which makes the
spin-conserving intervalley scattering rate ${\tau^{{\rm K}\rightarrow{\rm
      K'}}_{p~\Downarrow}}$ of spin-down holes faster than that
${\tau^{{\rm K'}\rightarrow{\rm K}}_{p~\Uparrow}}$ of spin-up
holes (see Fig.~\ref{figyw1}). Specifically, due to the large difference of
Fermi energies of spin-up and 
-down holes, the Rashba SOC [Eq.~(\ref{Rashba})] near the Fermi energy of
spin-up holes transfers holes from the spin-up states into the spin-down ones
with the same energies, and hence more spin-down
holes occupying the states with the energies higher than the corresponding Fermi
energy, leading to the larger effective hot-hole temperature. According
to our calculation, with the experimental obtainable hole density and the spin
polarization reaching $60~\%$, the accessible valley polarization can reach
beyond $1~\%$ which can last hundreds of picoseconds. In addition, due to
the absence of electrons in the conduction bands in $p$-type BL WSe$_2$, the
valley-depolarization induced by the exchange interaction\cite{depolarization}
is absent here. Therefore, the valley polarization proposed above can be
measured experimentally. This has not yet been reported in the literature.

This paper is organized as follows. In Sec.~{\ref{model}}, we introduce our
model and lay out the KSBEs. Then in Sec.~{\ref{Results_A}}, we investigate the
temperature and the hole density dependence of the out-of-plane spin
relaxation in BL WSe$_{2}$ with weak HF effective magnetic field. The anisotropy
of the spin relaxation is also presented in this part. In
Sec.~{\ref{Results_B}}, we show the influence of HF effective magnetic field
when the large out-of-plane spin polarization is considered. The valley
polarization due to the large spin polarization is also addressed in this
part. We summarize in Sec.~{\ref{summary}}.

\section{MODEL and KSBEs} 
\label{model}

In BL TMDs, in the presence of an external perpendicular electric field,
the effective Hamiltonian of the lowest two hole bands near the K
(K$'$) point can be written as\cite{electric2}
\begin{equation}
H_{\rm eff}^{\mu}=\varepsilon_{\mu{\bf k}}+{\bf \Omega}^{\mu}\cdot{\bm \sigma}/2,
\end{equation}
where $\varepsilon_{\mu{\bf k}}={\bf k}^2/(2m^{*})$, with $m^*$ being the
effective mass of the hole; the Rashba SOC ${\bf \Omega}^{\mu}$ is given
in Eq.~({\ref{Rashba}}). However, the Rashba coefficients $\nu$, $\alpha$ and
$\eta$ are absent in the literature. 
In this work, we use
the ML Rashba SOC\cite{R} in each layer to construct the BL Rashba
SOC. Specifically, a hole in the lowest hole band in a given layer 
can first hop into the second layer by the spin-conserving interlayer
hopping and then flip spin by the ML Rashba SOC to the lowest hole band in the
second layer. Or the hole in the lowest hole band can first flip spin in the
first layer and then hop to the second layer with same spin. The
lowest two hole bands in BL TMDs, which have opposite spin
polarizations, are then coupled with each other. Due to the large spin splitting of the hole
bands in each layer,  both the ML Rashba SOC and interlayer hopping can be
treated perturbatively. We use the L{\"o}wdin partition
method\cite{dia_1,dia_2} to derive the effective Rashba SOC of the lowest two
hole bands in BL TMDs, and obtain the Rashba coefficients, which are given in
Appendix~\ref{AA} in details.

The KSBEs are then written as\cite{HF_2,pw_a,pw_b}
\begin{equation}
\partial_{t}\rho_{\mu {\bf k}}=\partial_{t}\rho_{\mu {\bf k}}|_{\rm
  coh}+\partial_{t}\rho_{\mu {\bf k}}|_{\rm scat},
\end{equation}
where $\rho_{\mu {\bf k}}$ represent the density matrices of holes with
the off-diagonal terms $\rho_{\mu {\bf
    k},\frac{1}{2}-\frac{1}{2}}=\rho^*_{\mu {\bf
    k},-\frac{1}{2}\frac{1}{2}}$ representing the spin coherence and the
diagonal ones $\rho_{\mu {\bf
    k},\sigma\sigma}{\equiv}f_{\mu{\bf k},{\sigma}}~~({\sigma}=\pm\frac{1}{2})$ being the hole
distribution functions. The coherent terms $\partial_{t}\rho_{\mu {\bf k}}|_{\rm
  coh}$ describe the spin precessions of holes due to the effective magnetic
field ${\bf \Omega}^{\mu}$ and the HF effective magnetic field ${\bf
  \Omega}_{\rm HF}$. The scattering terms $\partial_{t}\rho_{\mu {\bf k}}|_{\rm
  scat}$ include the hole-hole Coulomb, long-range hole-impurity, intravalley
hole-phonon and especially the intervalley hole-phonon scattering. The detailed
expressions for the coherent and scattering terms can be found in
Ref.~\onlinecite{pw_a}. Particularly, the large Zeeman-like
term $\mu\eta{E_{z}}$ in the Rashba SOC [Eq.~(\ref{Rashba})], leads to the
energy-splitting of the lowest two hole bands in BL TMDs, which has opposite
signs in the two valleys, and hence affects the spin-conserving  
intervalley hole-phonon scattering but has no effect on the intravalley
scattering. By considering this energy-splitting,
the intervalley scattering parts in the KSBEs are given by\cite{HF_2} 
\begin{eqnarray}
\partial_{t}\rho_{\mu {\bf k}}|_{\rm scat}^{\rm inter}=\big\{S_{\mu{\bf
    k}}(<,>)+S_{\mu{\bf k}}(<,>)^{\dagger}\big\}-\{<\leftrightarrow>\},~~~~ \label{interscatt}
\end{eqnarray}
with
\begin{eqnarray}
\nonumber
S_{\mu{\bf k}}(>,<)=\pi{\sum_{{\bf
      q}\mu'\eta_1\eta_2}}|{M^{\lambda}_{\mu\mu'{\bf q}}}|^2\rho^{>}_{\mu'{\bf 
    k-q}}T_{\mu'\eta_1}T_{\mu\eta_2}\rho^{<}_{\mu{\bf k}}\big[N^{<}_{\bf
  q}&&\nonumber\\
\times\delta(\varepsilon_{\eta_2{\bf k}}-\varepsilon_{\eta_1{\bf k-q}}+\omega_{\bf
  q})+N^{>}_{\bf q}\delta(\varepsilon_{\eta_1{\bf k-q}}-\varepsilon_{\eta_2{\bf
    k}}+\omega_{\bf q})\big].&&~
\end{eqnarray}
Here, $N^{<}_{\bf q}=N_{\bf q}$ is the phonon distribution, $N^{>}_{\bf
  q}=N_{\bf q}+1$; $\rho^{<}_{\mu{\bf k}}=\rho_{\mu{\bf k}}$, and
$\rho^{>}_{\mu{\bf k}}=1-\rho_{\mu{\bf k}}$; $\eta_1 (\eta_2)=\pm{1}$ and
$\varepsilon_{\pm{1}{\bf k}}=\varepsilon_{\bf k}\pm\eta{E_{z}}$; The projector
operator reads $T_{\mu\eta}=(1+\eta\mu\sigma_{z})/2$;
$|{M^{\lambda}_{\mu\mu'{\bf q}}}|^2$ is the scattering matrix element of the
intervalley phonon mode $\lambda$.

However, the hole-phonon scattering matrix elements in BL TMDs
have not yet been reported in the literature. 
For the intervalley hole-phonon
scattering, with the intervalley scattering in each layer suppressed due to the
large spin splitting,\cite{s_1}  holes in a given valley at a given layer are
scattered into the other valley at different layer.  We derive the matrix
elements of this intervalley scattering by using the tight-binding model
according to the arXiv version of the work by Viljas and
Heikkil{\"a}.\cite{ph_a} We show that only the phonon modes $K_{6}^{\rm L}$ and
$K_{6}^{\rm H}$ make the main contribution. Here, $K_{6}^{\rm L}$ ($K_{6}^{\rm
  H}$) is the phonon 
mode at the K point corresponding to the irreducible
representation $E^{\prime\prime}_{2}$ of group
$C_{3h}$ with the lower (higher) phonon energy.\cite{s_1}  
For the intravalley hole-phonon scattering, we derive the matrix elements of
out-of-plane phonon according to the same work by Viljas and
Heikkil{\"a}; the matrix elements of in-plane phonon in BL TMDs are constructed by using
the ML ones, which have been reported in the work by Jin {\em et
  al}..\cite{ph_6} The specific derivation of the hole-phonon scattering matrix
elements is give in Appendix~\ref{BB}.

The matrix elements of the hole-phonon scatterings including the intervalley
$K_{6}^{\rm L}$ ($|{M^{\rm K_{6}^{L}}_{\mu\mu'{\bf q}}}|^2$) and $K_{6}^{\rm H}$
($|{M^{\rm K_{6}^{H}}_{\mu\mu'{\bf q}}}|^2$) phonon scatterings and the
intravalley in-plane acoustic (AC) ($|{M^{\rm AC}_{\mu\mu'{\bf q}}}|^2$),
in-plane optical (OP) ($|{M^{\rm E^{2}_{2g}}_{\mu\mu'{\bf q}}}|^2$,
$|{M^{\rm E_{1u},E^{1}_{2g}}_{\mu\mu'{\bf q}}}|^2$) and out-of-plane OP
($|{M^{\rm B^{2}_{2g}}_{\mu\mu'{\bf q}}}|^2$, $|{M^{\rm
    A^2_{2u},B^{1}_{2g}}_{\mu\mu'{\bf q}}}|^2$) phonon scatterings are given by  
\begin{eqnarray} 
&&|{M^{K_{6}^{\rm L}}_{\mu\mu'{\bf
      q}}}|^2=\frac{t^{\prime}_{\perp}}{2\rho\Omega_{{\rm K},K_{6}^{\rm L}}}\delta_{\mu',-\mu},\\ 
&&|{M^{K_{6}^{\rm H}}_{\mu\mu'{\bf
      q}}}|^2=\frac{t^{\prime}_{\perp}}{2\rho\Omega_{{\rm K},K_{6}^{\rm
      L}}}\frac{2M_{\rm M}}{M_{\rm X}}\delta_{\mu',-\mu}, \\
&&|{M^{{\rm AC},{E^{2}_{\rm 2g}}}_{\mu\mu'{\bf
      q}}}|^2=\frac{(\Xi)^2q}{2{\rho}v_{\rm
    LA}}\delta_{\mu',\mu},\\ 
&&|{M^{B^{2}_{\rm 2g}}_{\mu\mu'{\bf
      q}}}|^2=\frac{{t^{\prime}_{\perp}}^2}{{\rho}{\gamma}q^2}\delta_{\mu',\mu},\\      
&&|{M^{E_{\rm 1u},E^{1}_{\rm 2g}}_{\mu\mu'{\bf
      q}}}|^2=\frac{(D_{\rm OP})^2}{{\rho}\Omega_{\Gamma,{
      E_{\rm 1u}}}}\delta_{\mu',\mu},\\
&&|{M^{A^2_{\rm 2u},B^{1}_{\rm 2g}}_{\mu\mu'{\bf
      q}}}|^2=\frac{{t^{\prime}_{\perp}}^2}{2{\rho}\Omega_{\Gamma,{ 
      A^2_{\rm 2u}}}}\frac{2M_d}{M_t}\delta_{\mu',\mu},
\end{eqnarray}
where  $E^{2}_{\rm 2g}$,
$B^{2}_{\rm 2g}$, $E_{\rm 1u}$, $E^{1}_{\rm 2g}$, $A^2_{\rm 2u}$, 
$B^{1}_{\rm 2g}$ are the phonon modes at the $\Gamma$ point in BL
TMDs;\cite{ph_v1,ph_v2,ph_v3,ph_v4} $\Omega_{{\rm
    \Gamma (K)},\lambda}$ is the energy of phonon mode $\lambda$ at ${\rm
  \Gamma (K)}$ point; $\rho$ is the
mass density; $v_{\rm LA}$ is the sound velocity
corresponding to longitudinal acoustic (LA) phonon;
$\gamma=\sqrt{k_BT/({\rho}v_0^2)}$ is the parameter for the energy dispersion of
the out-of-plane AC phonon with $v_0$ being the corresponding sound
velocity;\cite{s_1} $k_B$ represents the Boltzmann constant; $t^{\prime}_{\perp}$ is the 
derivative of interlayer hopping $t_{\perp}$ with respect to the corresponding
hopping bond length; $\Xi$ and $D_{\rm OP}$ are the deformation potentials of
the in-plane AC and OP phonon in ML TMDs,\cite{ph_6} respectively;  $M_{\rm M}$  
and $M_{\rm X}$ are the masses of the M atom and the X atom of the TMD MX$_2$, respectively.
The remaining matrix elements of the hole-hole Coulomb and long-range
hole-impurity scatterings can be found in the work by Wang and
Wu.\cite{s_4}

\section{NUMERICAL RESULTS} 

\begin{table}[htb]
  \caption{Parameters for WSe$_2$ used in our calculation. Note that $m_0$ stands for the
    free electron mass. } 
  \label{parameter} 
  \begin{tabular}{l l l l}
    \hline
    \hline
    $\nu~$({\r A}$^2$)&\quad$0.0342^a$&\quad\quad$\eta~$({\r A})&\quad$1.06^a$\\
    $m^*/m_0$&\quad$0.51^b$&\quad\quad$\alpha~$({\r
      A})$^2$&~$-5.673^a$\\ 
    $\kappa$&\quad$5.20^c$&\quad\quad$t^{\prime}_{\perp}~$(eV/{\r A})&\quad$1.26^d$\\ 
    $\Xi~$(eV)&\quad$2.1^e$&\quad\quad$D_{\rm OP}~$($10^8~$eV/cm)&\quad$3.1^e$\\
    $v_{\rm LA}~$($10^5~$cm/s)&\quad$3.30^e$&\quad\quad$v_0~$($10^5~$cm/s)&\quad$2.24^c$\\ 
    $\Omega_{E_{\rm 1u}}~$(meV)&\quad$30.8^e$&\quad\quad$\Omega_{A^2_{\rm 2u}}~$(meV)&\quad$38.51^f$\\
    $\Omega_{{\rm K},K_{6}^{\rm L}}~$(meV)&\quad$17.5^e$&\quad\quad$\Omega_{{\rm
      K},K_{6}^{H}}~$(meV)&\quad$31.5^e$\\
  $M_{\rm Se}/M_{\rm
    W}$&\quad$0.429$&\quad\quad$\rho~$($10^{-7}~$g/cm$^{2}$)&\quad$6.6^c$\\
    \hline
    \hline
\end{tabular}\\
 $^a$ Appendix A. \quad$^b$ Ref.~\onlinecite{Hamiltonian1}. \quad$^c$ Ref.~\onlinecite{s_1}. \quad$^d$
 Ref.~\onlinecite{inter-layer}. \\\quad$^e$ Ref.~\onlinecite{ph_6}. \quad$^f$
 Ref.~\onlinecite{ph_v4}. 
\end{table}

In this work, we focused on the case of WSe$_2$. The initial occupations of
holes of each spin are identical in the two valleys in our calculation. The
spin-polarization direction is along the ${\hat z}$-axis unless otherwise 
specified. The long-range impurity density is taken to be $N_{i}=0.02N_{h}$ with
$N_h$ being the hole density. Then the corresponding mobility at $T=250~$K in
our investigation is around $180~$cm$^2$$/$(V$\cdot$s), which agrees with those reported
in the existing experiments.\cite{m_1,m_2,m_3} All the material parameters 
used in our calculation are listed in Table~\ref{parameter}. With above
parameters, by numerically 
solving the KSBEs, we discuss the spin relaxation 
at small and large spin polarizations (i.e., weak and
strong HF effective magnetic fields) in Secs.~\ref{Results_A}
and~\ref{Results_B}, respectively.    

\subsection{Weak HF effective magnetic field}
\label{Results_A}
In this subsection, the initial spin polarizations $P_s$ in the two valleys are
set to be $2.5~\%$,  leading to the weak HF effective magnetic field
($|\Omega_{\rm HF}|{\ll}|\eta{E_{z}}|$) in our
calculation. Then the out-of-plane SRTs $\tau_{sz}^{\mu}$ due to the intravalley
scatterings [given in Eq.~(\ref{tau_s}) with $\Omega^{\mu}_{\rm
  eff}\approx\mu{\eta}E_{z}$], have 
identical values $\tau_{sz}$ in the two valleys. 
Moreover, owing to the competition between 
$[|\Omega^{\mu}_{\rm eff}|^2/\langle\Omega_{\perp}^2({\bf k})\rangle]\tau_p$
and $[\langle\Omega^2_{\perp}({\bf k})\rangle\tau_{p}]^{-1}$ in
Eq.~(\ref{tau_s}), the SRT can be
divided into two regimes: regime I, 
the anomalous EY-like regime, where the first term makes a more important
contribution and hence the SRT shows the EY-like behavior; regime II, the normal
strong scattering regime, where the second term becomes more important. The
crossover between regimes I and II is determined by $|\Omega_{\rm eff}\tau_p|{=}1$
with the system sitting in regime I (II) when $|\Omega_{\rm eff}\tau_p|>(<)1$.
In this work, with $|\Omega_{\rm eff}|\approx|{\eta}E_z|$ at small spin
polarization, we discuss the temperature and hole density dependence of the
out-of-plane SRT at the small ($|{\eta}E_{z}\tau_p|{\ll}1$) and medium 
($|{\eta}E_{z}\tau_p|{\approx}1$) fields. The large field with
$|{\eta}E_{z}\tau_p|{\gg}1$, which leads to a large energy-splitting of the  
hole bands similar to ML WSe$_{2}$, is not included in this work. We also
investigate the anisotropic spin relaxation by varying the spin-polarization
direction in this subsection.     
 
\subsubsection{Temperature dependence of out-of-plane spin relaxation}

\begin{figure}[htb]
  {\includegraphics[width=8.4cm]{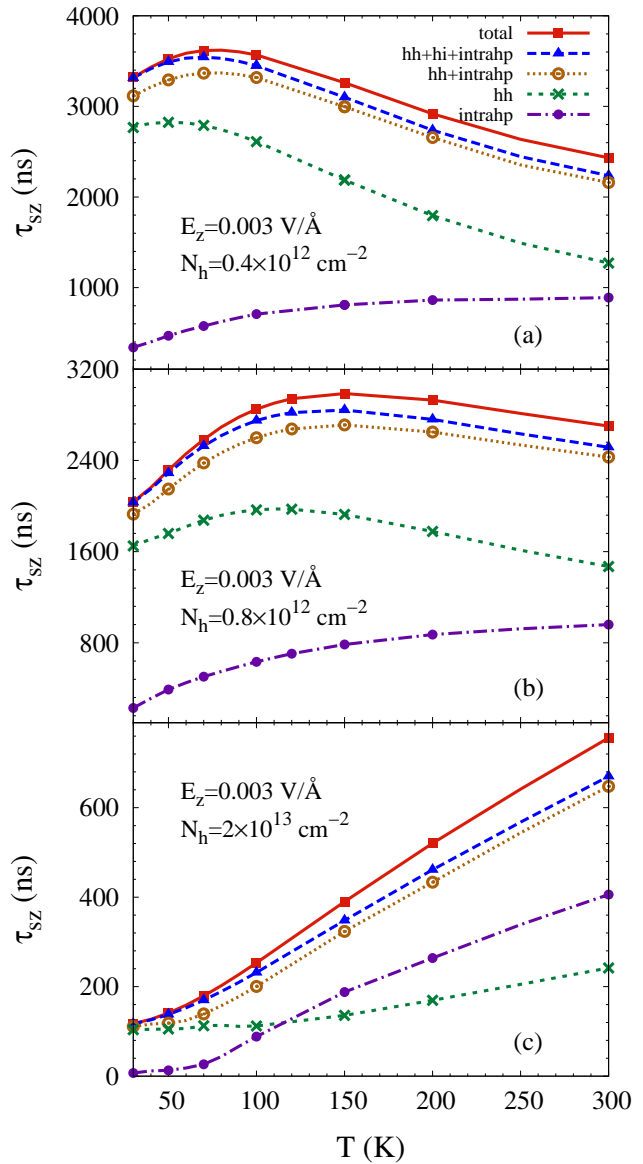}}
 \caption{(Color online) The SRTs $\tau_{sz}$ as function of $T$ at (a)
   $N_h=0.4{\times}10^{12}~$cm$^{-2}$, (b) $N_h=0.8{\times}10^{12}~$cm$^{-2}$,
   and (c) $N_h=2{\times}10^{13}~$cm$^{-2}$.  Squares: all the relevant
   scatterings are included; Triangles: the intervalley hole-phonon scattering 
   is removed; Open circles: both the intervalley hole-phonon and long-range
   hole-impurity scatterings are removed; Crosses (Dots): only the
   hole-hole Coulomb 
   (intravalley hole-phonon) scattering is included.
   $E_z=0.003~$V/\r{A}. }   
\label{figyw2}
\end{figure}

\begin{figure}[htb]
  {\includegraphics[width=8.2cm]{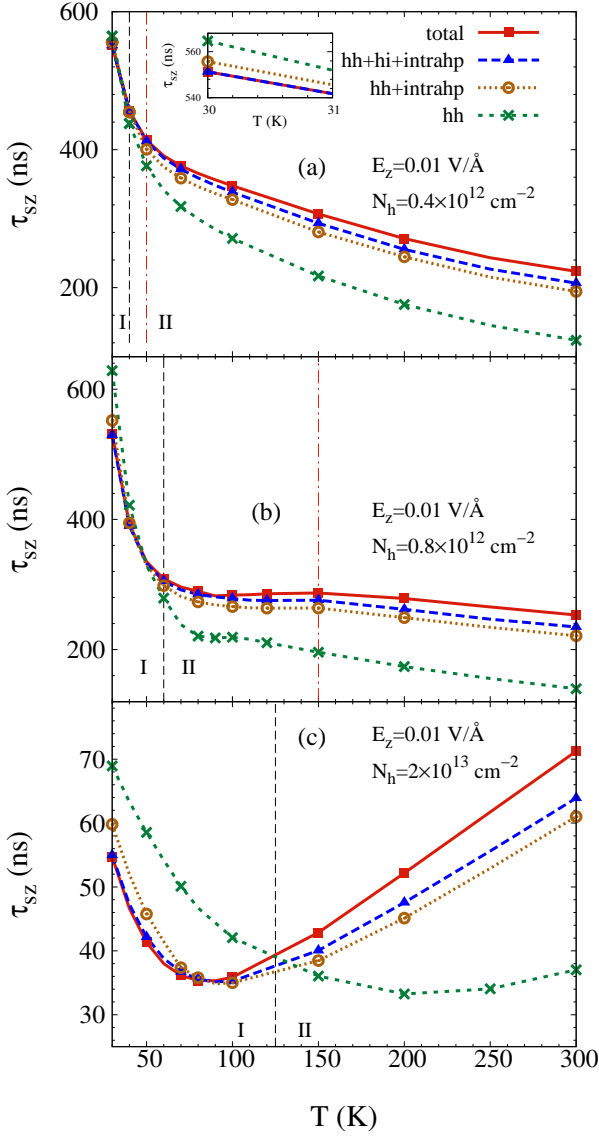}}
 \caption{(Color online) The SRTs $\tau_{sz}$ as function of $T$ at (a)
   $N_h=0.4{\times}10^{12}~$cm$^{-2}$, (b) $N_h=0.8{\times}10^{12}~$cm$^{-2}$,
   and (c) $N_h=2{\times}10^{13}~$cm$^{-2}$. Squares: all the relevant
   scatterings are included; Triangles: the intervalley hole-phonon scattering 
   is removed; Open circles: both the intervalley hole-phonon and long-range
   hole-impurity scatterings are removed; Crosses: only the
   hole-hole Coulomb scattering is included. The inset in (a) zooms the
   temperature range 
       $30$-$31$~K.  
   The vertical black dashed and red dash-dot 
   lines indicate $T^r_c$ and $T^{h}_c$, respectively. $E_z=0.01~$V/\r{A}.} 
\label{figyw3}
\end{figure}

We first investigate the temperature dependence of the out-of-plane spin
relaxation. The SRTs $\tau_{sz}$ as function of temperature $T$ at different
hole densities are plotted in Figs.~\ref{figyw2} and~{\ref{figyw3}} at the
small ($E_{z}=0.003~$V/\r{A}) and medium ($E_{z}=0.01~$V/\r{A}) fields,
respectively. All the relevant scatterings are included in the calculation. From
Fig.~\ref{figyw2}, it is noted that the intervalley hole-phonon and the
long-range 
hole-impurity scatterings are marginal to the out-of-plane spin relaxation,
since the SRTs with both the
intervalley and long-range hole-impurity scatterings (curve with open circles) 
removed are close to the one with all scatterings included (curve with
squares). To further elucidate the role of the remaining hole-hole Coulomb and
intravalley 
hole-phonon scatterings, we compare the SRT with only the hole-hole Coulomb
scattering included (curve with crosses) and the one with only the
intravalley hole-phonon scattering included (curve with dots). The
hole-hole Coulomb scattering is found to be more important at low temperature
since the SRT with only the hole-hole Coulomb scattering is much
larger than the counterpart, while the intravalley hole-phonon scattering
plays an important role at high temperature.  


At the small field (the SRT is in regime
II), a peak around $T=60~$K at 
the hole density 
$N_{h}=0.4{\times}10^{12}~{\rm cm}^{-2}$ (the Fermi temperature $T_F=11~$K) is
observed in the 
temperature dependence of the SRT [see Fig.~\ref{figyw2}(a)]. As revealed in the previous
works,\cite{pw_a,pw_2,pw_4} this peak is due to the crossover of the
hole-hole Coulomb scattering from the degenerate to 
nondegenerate limits [i.e., $1/\tau^{\rm
  hh}_{p}{\propto}\ln(T_{F}/T)T^{2}/T_{F}$ when $T{\ll}T_{F}$ and
$1/\tau^{\rm hh}_{p}{\propto}1/T$ when $T{\gg}T_{F}$]\cite{pw_3,pw_3.0} when
the SRT is in the normal strong scattering regime. Unlike the previous 
studies in semiconductors, where the peak locations in different materials
locate in the range $(T_F/4,2T_F)$,\cite{pw_a,pw_2,pw_4,pw_5,pw_6,pw_8} 
the peak location of $p$-type WSe$_2$ is around $5T_F$. Moreover,
compared with the case with only the hole-hole Coulomb scattering, 
with all scatterings included, the peak is shifted toward a higher
temperature at $T^{h}_c$ for the low density and found to be destroyed
(not shown) when
$N_{h}{\gtrsim}2{\times}10^{12}~{\rm cm}^{-2}$ ($5T_F{\gtrsim}275~$K). This is
due to the fact that the reduction of the Coulomb scattering in the
nondegenerate limit can
be compensated by the increase of the intravalley hole-phonon
scattering, similar to the previous work,\cite{pw_a} and
the intravally hole-phonon scattering is stronger than the hole-hole Coulomb
scattering after $275~$K. Consequently, when
$N_{h}{\gtrsim}2{\times}10^{12}~{\rm cm}^{-2}$, the SRT increases with 
the decrease of $\tau_p$ by increasing temperature, as shown in
Fig.~\ref{figyw2}(c) at $N_{h}=2{\times}10^{13}~{\rm cm}^{-2}$.

At the medium field (see Fig.~\ref{figyw3}), the crossover between 
regimes I and II is observed at $T^r_c$. Specifically, with the increase of
$\tau_p$ by 
removing each scattering at the same temperature, $\tau_{sz}$ increases in
regime I while decreases in regime II. Compared with the location of $T^h_c$
at the small field, which is determined by $\tau_p$
and not influenced by $E_z$, we find that the location of the crossover between
regimes I and II 
$T^r_c$ (indicated by the vertical black dashed line) is always smaller than
$T^h_c$ (indicated by the vertical red dash-dot line) at the corresponding hole
density. This is because that with 
increasing temperature before $T^h_c$, the decrease of $\tau_p$ by the
hole-hole Coulomb and intravalley hole-phonon scatterings leads to the crossover
from regime I ($|{\eta}E_{z}\tau_p|>1$) to II
($|{\eta}E_{z}\tau_p|<1$). Therefore, with
increasing temperature before $T^h_c$, the SRT in regime I
($\tau_{sz}\propto\tau_p$) decreases whereas in regime II
($\tau_{sz}\propto\tau_p^{-1}$) increases [see Fig.~\ref{figyw3}(b)]. Additionally, with increasing
temperature after $T^h_c$ (the SRT is entirely in regime II), as mentioned
above, $\tau_p$ increases (decreases) when 
$N_{h}{<(>)}2{\times}10^{12}~{\rm cm}^{-2}$, and hence 
$\tau_{sz}$ decreases (increases), as shown in the figures. Consequently,
when $T^r_c$ is close to 
$T^h_c$, as shown in Fig.~\ref{figyw3}(a) at $N_{h}=0.4{\times}10^{12}~{\rm
  cm}^{-2}$, the
peak existed at the previous small field
with the same hole density is 
vanished here. But when $T^r_c$ is separated from $T^h_c$, as shown in
Fig.~\ref{figyw3}(b) at $N_{h}=0.8{\times}10^{12}~{\rm
  cm}^{-2}$,  the peak existed at the small field with the identical hole 
density [Fig.~\ref{figyw2}(b)] is observed at the same temperature
    position here, but is less visible.

\subsubsection{Hole density dependence of out-of-plane spin relaxation} 

\begin{figure}[htb]
  {\includegraphics[width=8.5cm]{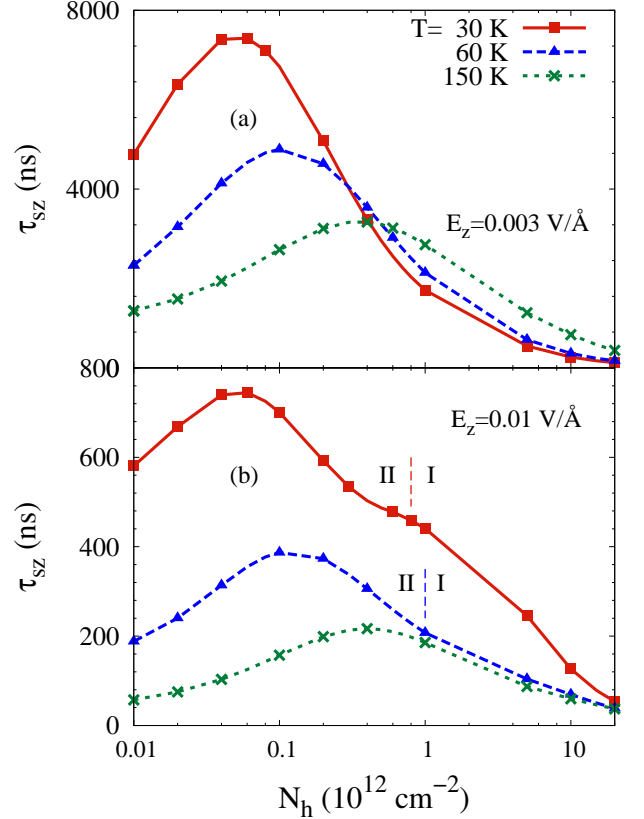}}
 \caption{(Color online) The SRTs $\tau_{sz}$
   versus $N_h$ at different temperatures with (a)
   $E_z=0.003~$V/\r{A} and (b) $E_z=0.01~$V/\r{A}. The vertical red (blue) dashed  
   line indicates the crossover between regimes I and II at $T=30~(60)~$K. }
\label{figyw4}
\end{figure}

Next we turn to study the hole density dependence of out-of-plane spin
relaxation. The SRTs $\tau_{sz}$ versus the hole density $N_{h}$ at
different temperatures are plotted in Figs.~\ref{figyw4}(a) and (b) at the small
($E_{z}=0.003~$V/\r{A}) and medium ($E_{z}=0.01~$V/\r{A}) fields, respectively. 
At the small field, a peak around
$T_{F}{\approx}T/4.5$ is observed in the
density dependence of the SRT, which is due to the
crossover of holes from the nondegenerate to degenerate limits when
the SRT is in the normal strong scattering regime.\cite{pw_4,pw_5,pw_6,pw_8}  
At the medium field, similar peak is also observed, as shown in
Fig.~\ref{figyw4}(b), indicating the SRT around $T_{F}{\approx}T/4.5$ still
sits in regime II.  But compared with the results at the small field, 
it is noted that the decrease of the SRT in the degenerate
limit at the medium field slows down after $N_{h}=0.8{\times}10^{12}{\rm
  cm}^{-2}$ at $T=30~$K (curve with squares). This arises from the crossover between regimes I and II.
Specifically, with increasing the hole density in the degenerate limit at low
temperature, the increase of $|{\eta}E_{z}\tau_{p}|$ due to the dominant
hole-hole Coulomb scattering
[i.e., $1/\tau^{\rm hh}_{p}{\propto}\ln(T_F/T)T^2/T_F$ when 
$T_{F}{\gg}T$ ($T_F{\propto}N_h$ for the two-dimensional case)]
drives the SRT from regime II ($|{\eta}E_{z}\tau_p|<1$) to regime
I ($|{\eta}E_{z}\tau_p|>1$).  In the degenerate limit,
one has $\tau_{sz}{\propto}\ln(T_F/T)T^2/T_F^2$ in regime II whereas
$\tau_{sz}{\propto}1/[T^2\ln(T_F/T)]$ in regime I since
    $\langle\Omega_{\perp}^2({\bf k})\rangle{\propto}T_F$ in
    Eq.~(\ref{tau_s}). Consequently, with the increase of $N_h$, the
decrease of SRT slows down after the 
crossover. Similar behavior can also be observed at $T=60~$K (curve with
    triangles).
But at high temperature $T=150~$K (curve with crosses), the intravalley
hole-phonon scattering becomes stronger and the reduction of the Coulomb
scattering in the degenerate limit can be compensated by the increase of the
hole-phonon scattering, leading to the crossover indistinguishable.

\subsubsection{Anisotropic  spin relaxation}

\begin{figure}[htb]
  {\includegraphics[width=8.5cm]{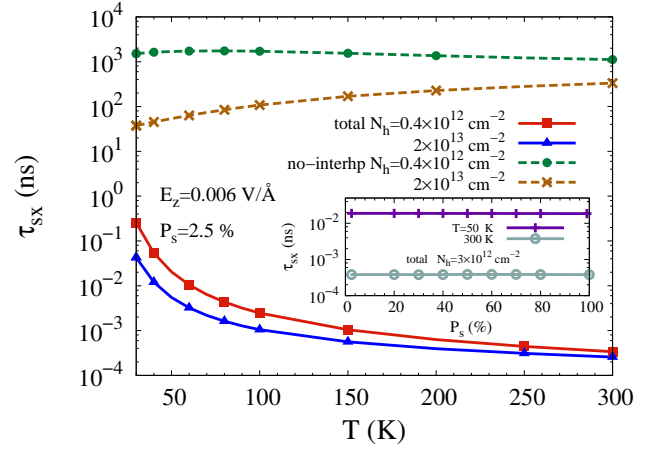}}
  \caption{(Color online) The SRTs $\tau_{sx}$ as function of $T$ at different
    hole densities. $P_s=2.5~\%$. Solid curves: all
    the relevant 
    scatterings are included; Dashed curves: the intervalley hole-phonon
    scattering is removed.  The inset shows the SRTs versus
    initial spin polarization $P_s$ at different temperatures when
    $N_h=3{\times}10^{12}~$cm$^{-2}$. $E_z=0.006~$V/\r{A}. 
   }
\label{figyw5}
\end{figure}

\begin{figure}[htb]
  {\includegraphics[width=8.5cm]{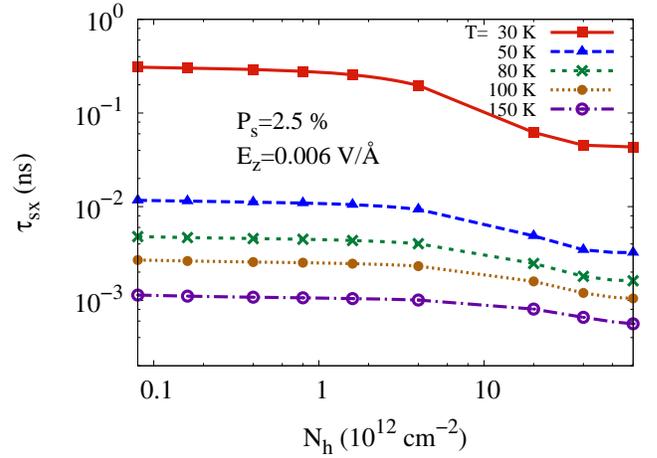}}
  \caption{(Color online) The SRT $\tau_{sx}$ versus hole density $N_h$ at
    different temperatures.
    $P_s=2.5~\%$. $E_z=0.006~$V/\r{A}. }   
  
\label{figyw6}
\end{figure}

We also address the anisotropy of the spin relaxation with respect to the spin 
polarization direction. The temperature and hole density dependences 
of the in-plane spin relaxation along the ${\hat x}$-axis are plotted in
Figs.~\ref{figyw5} and~{\ref{figyw6}}, respectively. From Fig.~\ref{figyw5},
at the same hole density and temperature,
the SRT with all the relevant scatterings included (solid curves with squares
or triangles) is about four orders of magnitude smaller than the one with the
intervalley hole-phonon scattering removed (dashed curves with dots or
crosses), indicating the
intervalley hole-phonon scattering makes a dominant contribution to the
in-plane spin relaxation. This is very different from the previous
results of the out-of-plane spin relaxation, where the intervalley hole-phonon
scattering is marginal. Specifically, the Zeeman-like term, together 
with the intervalley hole-phonon scattering, opens an intervalley spin relaxation
channel, as pointed out by Wang and Wu in ML 
MoS$_2$,\cite{s_4,s_5} which dominates the in-plane spin relaxation in
$p$-type BL WSe$_2$.

In addition, the SRTs decrease
monotonically with the increase of temperature (see Fig.~\ref{figyw5})
or hole density (see Fig.~\ref{figyw6}) but are insensitive to the initial spin
polarization (see the inset in Fig.~\ref{figyw5}). 
This is due to the dominant intervalley spin relaxation channel, very similar to the case of
ML MoS$_2$\cite{s_4} when only the intervalley electron-phonon scattering is considered.
Specifically, the intervalley hole-phonon scattering is 
in the weak-scattering limit, and hence the in-plane SRT
$\tau_{sx}=\tau^{\rm inter}_p$,\cite{s_4,pw_9} with $\tau^{\rm inter}_p$
representing the 
intervalley hole-phonon scattering time. Consequently, the in-plane
SRT decreases with the enhancement of the intervalley hole-phonon
scattering as the temperature or hole density increases.

\subsection{Strong HF effective magnetic field}
\label{Results_B}

\subsubsection{Spin polarization dependence of out-of-plane spin relaxation}

\begin{figure}[htb]
  {\includegraphics[width=8.4cm]{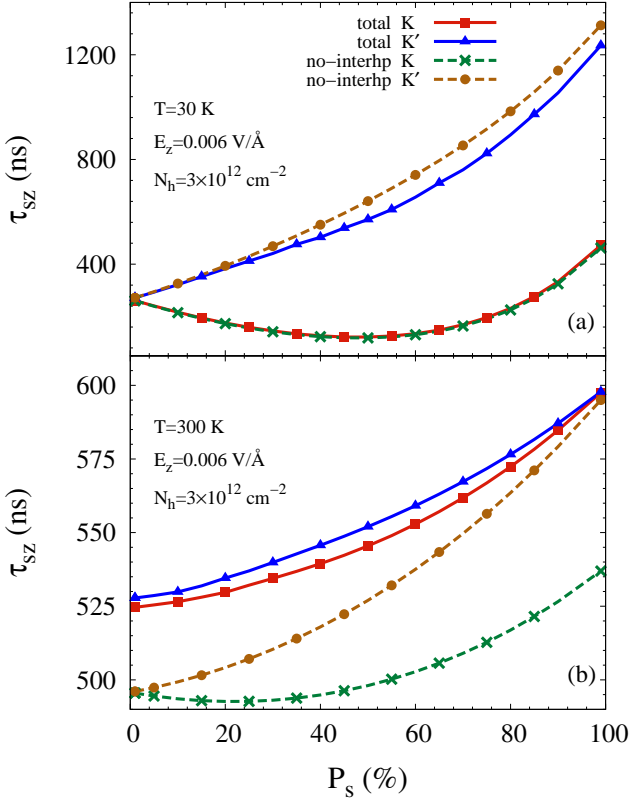}}
  \caption{(Color online) The SRTs $\tau_{sz}$ versus initial spin
    polarization $P_s$ at (a) $T=30~$K and (b) $T=300~$K. Squares (Triangles): in the K (K$'$) valley with all
    the relevant scatterings included; Crosses (Dots): in the K (K$'$) valley
    with the intervalley  hole-phonon scattering
    removed. $N_h=3{\times}10^{12}~$cm$^{-2}$. $E_z=0.006~$V/\r{A}.  } 
\label{figyw7}
\end{figure}

\begin{figure}[htb]
  {\includegraphics[width=8.4cm]{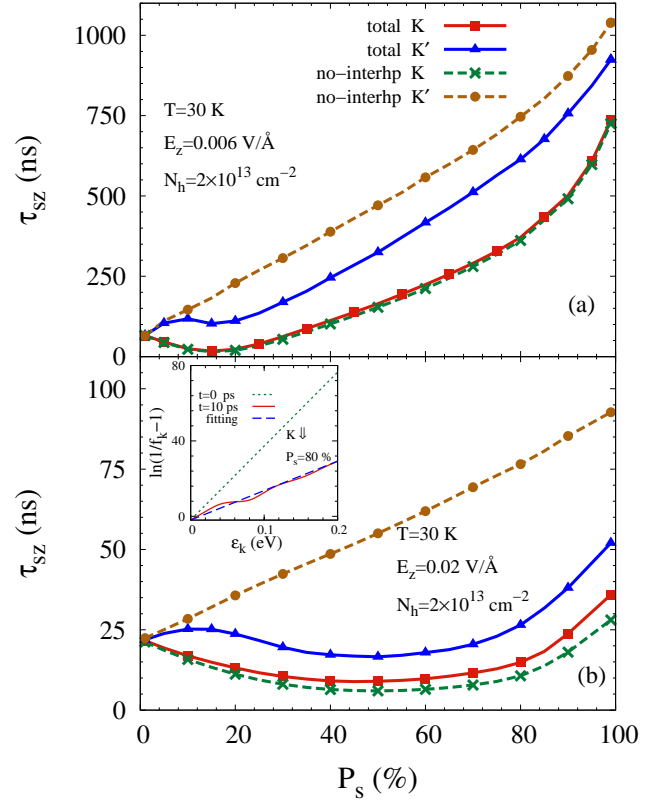}}
  \caption{(Color online)  The SRTs $\tau_{sz}$ versus initial spin
    polarization $P_s$ at (a)   $E_z=0.006~$V/\r{A} and (b) 
    $E_z=0.02~$V/\r{A}. Squares
    (Triangles): in the K (K$'$) valley with all the relevant scatterings
    included; Crosses (Dots): in the K (K$'$) valley with the intervalley
    hole-phonon scattering removed.  The inset in (b) shows the distributions
    for spin-down holes in the K valley at
    $N_h=2{\times}10^{13}~$cm$^{-2}$ and initial spin polarization $P_s=80~\%$
    when $t=0~$ps and $t=10~$ps. $T=30~$K. $N_h=2{\times}10^{13}~$cm$^{-2}$.  }
\label{figyw8}
\end{figure}

At large spin polarization (i.e., strong HF effective magnetic field), different
total effective magnetic fields [Eq.~(\ref{total})] in the two valleys show
up. The out-of-plane SRTs in the two valleys are plotted as function of the
initial spin polarization $P_s$ in Figs.~\ref{figyw7} and~{\ref{figyw8}} at
different temperatures and hole densities. We first focus on the case with 
the intervalley hole-phonon scattering removed (dashed curves with crosses or
dots), where the out-of-plane spins relax independently in the two valleys. As
seen from the figures, different SRTs in the two valleys are obtained, which
arise from the difference of the total effective magnetic fields in
Eq.~(\ref{tau_s}). Specifically, when the HF and Zeeman-like effective
magnetic fields share the same direction, say, in the K$'$ valley, the total
effective magnetic field $|{\bm \Omega}^{\rm K'}_{\rm eff}|=|\Omega_{\rm HF}|+|\eta{E_{z}}|$, and
hence the SRT $\tau^{\rm K'}_{sz}$ (dashed curve with dots), proportional
to $1+|{\bm \Omega}^{\rm K'}_{\rm eff}\tau_p|^2$ [see Eq.~(\ref{tau_s})], 
increases with increasing the spin polarization (i.e., HF effective
magnetic field).  Whereas in the K valley, which has opposite directions between
the HF and Zeeman-like fields, the total 
effective magnetic field reads $|{\bm \Omega}^{\rm K}_{\rm
  eff}|=||\eta{E_{z}}|-|\Omega_{\rm HF}||$, and hence the SRT  $\tau^{\rm
  K}_{sz}$ (dashed curve with crosses), proportional to $1+|{\bm
  \Omega}^{\rm K}_{\rm eff}\tau_p|^2$, exhibits a minimum at
$|\Omega_{\rm HF}|=|\eta{E_{z}}|$ in the spin
polarization dependence.  This minimum location in the spin polarization
dependence can be tuned by the temperature [see the comparison between
Figs.~\ref{figyw7}(a) and~(b)], hole density [see the comparison between
Fig.~\ref{figyw7}(a) and Fig.~\ref{figyw8}(a)] and electric field [see the comparison between
Figs.~\ref{figyw8}(a) and~(b)].  Additionally,
due to the larger effective magnetic 
field in the K$'$ valley in the present configuration, $\tau^{\rm K}_{sz}$ is
always faster than $\tau^{\rm K'}_{sz}$.        

When the intervalley hole-phonon scattering is included (solid curves with
squares or triangles), the above difference of the SRTs between the two valleys is
suppressed.
This is understood that the
intervalley scattering suppresses the difference of the hole density of each
spin between the two valleys.  At low hole density and low temperature,
the intervalley scattering is weak, leading to the marginal suppression on
the difference between $\tau^{\rm K}_{sz}$ and  $\tau^{\rm K'}_{sz}$, as
shown in Fig~\ref{figyw7}(a) at $T=30~$K and $N_h=3{\times}10^{12}~$cm$^{-2}$.
However, in contrast to this marginal suppression at
$T=30~$K, the suppression with the same hole density at high temperature is
markedly strong, as shown in Fig.~\ref{figyw7}(b) at $T=300~$K, where the SRTs
with the intervalley hole-phonon scattering included are nearly identical in the two
valleys. This arises from the enhanced intervalley hole-phonon scattering. 

Moreover, even at low temperature $T=30~$K, where the 
intervalley hole-phonon scattering by absorbing phonons is negligible, 
the suppression on the difference between $\tau^{\rm K}_{sz}$ and $\tau^{\rm
  K'}_{sz}$ becomes stronger with the increase of the hole density [see the
comparison between Fig.~\ref{figyw7}(a) and Fig.~\ref{figyw8}(a)].
This arises from the different spin relaxation processes in the two valleys. 
Specifically, as shown in Fig.~\ref{figyw1}, in the temporal evolution,
the faster spin relaxation in the K
valley makes the Fermi energy for spin-down (-up) holes
larger (smaller) than the corresponding one with the same spin in the K$'$
valley, and this difference of Fermi energies
[given in
Eq.~(\ref{differenceCC}) in Appendix~\ref{CC}]  ,  
\begin{eqnarray}
\label{difference}
{\Delta}E_F^{\Uparrow(\Downarrow)}={0.25P_sN_h/D_s}(e^{-t/{\tau^{\rm
      K'}_{sz}}}-e^{-t/{\tau^{\rm K}_{sz}}}),
\end{eqnarray}
with ${\Delta}E_F^{\Uparrow(\Downarrow)}=D_s(N^{\Uparrow (\Downarrow)}_{\rm K'
  (K)}-N^{\Uparrow (\Downarrow)}_{\rm K (K')})$ and $D_s$ representing the density of states, 
triggers the intervalley scattering of spin-down holes from the K to the K$'$ 
valley by emitting phonons. In this situation, the
intervalley scattering time of 
spin-down holes in the degenerate limit can be written as [given in
Eq.~(\ref{vpfk4}) in Appendix~\ref{DD}]  
\begin{eqnarray} 
\label{intertime}
\tau^{{\rm K (K')}\rightarrow{\rm
    K' (K)}}_{p~\Downarrow
  (\Uparrow)}&=&\tau^*_p(e^{\beta{{\Delta}E^{\Downarrow (\Uparrow)}}}-1),
\end{eqnarray}
for ${\Delta}E^{\Downarrow (\Uparrow)}>0$. Here,
${\Delta}E^{\Downarrow(\Uparrow)}=\Omega_{\xi}-{\Delta}E_F^{\Uparrow(\Downarrow)}$   
with $\Omega_{\xi}$ ($\xi=K_{6}^{\rm L},K_{6}^{\rm H}$) standing for the
intervalley phonon energy; $1/\tau_p^*=2|M^{\xi}|^2m^*$ with $|M^{\xi}|$
being the matrix element of the intervalley hole-phonon scattering.
Focused on a specific case at $P_s=60~\%$, at low hole density
$N_h=3{\times}10^{12}~$cm$^{-2}$ shown in 
Fig.~\ref{figyw7}(a), the difference of the 
Fermi energies between the two valleys 
(${\Delta}E_F^{\Uparrow(\Downarrow)}|^{\rm max}{\approx}3.3~$meV)
is much smaller than the intervalley phonon energy ($\Omega_{{\rm
    K},K^{\rm L}_6}=17.5~$meV), and hence the  
intervalley scattering by emitting phonons is blocked since
$\beta\Delta{E^{\Downarrow(\Uparrow)}}{\gg}1$.  Nevertheless, with increasing
the hole density to $N_h=2{\times}10^{13}~$cm$^{-2}$, this difference 
over time (${\Delta}E_F^{\Uparrow(\Downarrow)}|^{\rm
  max}{\approx}15~$meV) becomes closer to the intervalley phonon
energy, leading to the intervalley scattering by emitting phonons enhanced at
nonzero temperature.  

In addition, we find that the above suppression on the difference of the
    SRTs at low temperature can be further enhanced by increasing the 
electric field [see the comparison between Fig.~\ref{figyw8}(a) and~(b)], which arises from a quasi 
hot-hole distribution. Specifically, as shown in the inset of 
Fig.~\ref{figyw8}(b), where the  
distributions for spin-down holes in the K valley at $t=0~$ps (dash-dot curve)
and $t=10~$ps (solid curve) are plotted at the initial spin polarization
$P_s=80~\%$. Due to the weak intravalley 
hole-phonon scattering at low temperature but relatively strong hole-hole
Coulomb scattering,  one finds that the fast
spin precessions at a large electric field (i.e., a large Rashba SOC) result in
a quasi hot-hole Fermi distribution characterized by an effective hot-hole
temperature $T_{\rm eff}$ (the slope of the dashed curve is proportional to 
$T_{\rm eff}^{-1}$) larger than $T$ (the slope of the dash-dot curve is
proportional to $T^{-1}$), which enhances the intervalley hole-phonon 
scattering for spin-down holes by emitting 
phonons. Similar behaviors can also be observed for
spin-up holes. With the enhanced intervalley hole-phonon scattering, the
difference of the SRTs between the two valleys is further suppressed. 

\subsubsection{Temperature dependence of out-of-plane spin relaxation}  

\begin{figure}[htb]
  {\includegraphics[width=8.5cm]{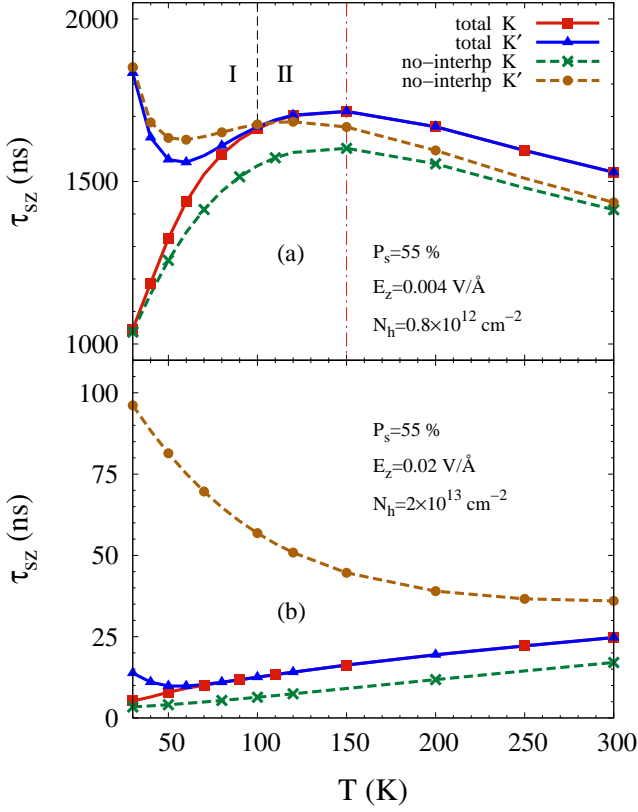}}
  \caption{(Color online) The SRT $\tau_{sz}$ versus temperature $T$ at (a) 
    $N_h=0.8{\times}10^{12}~$cm$^{-2}$ and  $E_z=0.004~$V/\r{A} and (b)
    $N_h=2{\times}10^{13}~$cm$^{-2}$ and $E_z=0.02~$V/\r{A}. The initial spin
    polarization $P_s=55~\%$. Squares (Triangles): in the K (K$'$) valley with
    all the relevant scatterings included; Crosses (Dots): in the K (K$'$)
    valley with the intervalley hole-phonon scattering removed. The vertical
    black dashed and red dash-dot  
   lines indicate $T^r_c$ and $T^{h}_c$, respectively.
}
\label{figyw9}
\end{figure}

The SRTs in the two valleys versus 
temperature $T$ at different hole densities are plotted in Fig.~\ref{figyw9}
with a fixed large spin polarization $P_s=55~\%$. In the computation, the electric
field is chosen at $|\eta{E_{z}}|=|\Omega_{\rm HF}|$ when $T=30~$K, providing a
large difference of the SRTs in the two valleys in the absence of the
intervalley hole-phonon scattering.   

At $N_h=0.8{\times}10^{12}~$cm$^{-2}$ shown in Fig.~\ref{figyw9}(a), for the case
with the intervalley hole-phonon scattering removed, even at the
small electric field ($E_{z}=0.004~$V/\r{A}), due
to the large HF effective magnetic field ($|\Omega_{\rm
  HF}|/\eta=0.0023~$V/\r{A} at $T=30~$K), 
the total effective magnetic field in the K$'$ valley ($|{
  \Omega}^{\rm K'}_{\rm eff}|=|\Omega_{\rm HF}|+|\eta{E_{z}}|$) is a medium
field ($|\Omega^{\rm K'}_{\rm eff}\tau_p|\approx1$) whereas that in the K valley
($|{\Omega}^{\rm K}_{\rm eff}|=||\Omega_{\rm HF}|-|\eta{E_{z}}||$) is a small field
($|\Omega^{\rm K'}_{\rm eff}\tau_p|\ll1$). Therefore, the behaviors of the SRTs originally existed
at small spin polarization with the small (medium) total effective magnetic
field [see Fig.~\ref{figyw2}(b) (\ref{figyw3}(b))] is observed in the 
K (K$'$) valley here [dashed curve with crosses (dots)].
In addition, the difference of the SRTs in the two valleys becomes weaker
    with the increase of temperature even when the intervalley hole-phonon
    scattering is removed.  
This is because the HF effective magnetic field becomes weaker with
increasing temperature into the nondegenerate limit,\cite{HF_1} making the difference of the total effective magnetic
fields between the two valleys much smaller. When the intervalley
hole-phonon scattering is included (solid curves with squares or triangles),
this difference of the SRTs is further suppressed and becomes nearly
vanished when $T>110~$K. 

At the high hole density 
$N_h=2{\times}10^{13}~$cm$^{-2}$ as shown in Fig.~\ref{figyw9}(b),
the HF effective magnetic field ($|\Omega_{\rm
  HF}|/\eta=0.022~$V/\r{A} at $T=30~$K) is insensitive to the temperature in
  the degenerate limit, and hence the SRT in the K$'$ valley with the
intervalley scattering removed is completely in regime I whereas that in the K
valley sits in regime II. Consequently, with the decrease of $\tau_p$ by
increasing temperature, the SRT increases in the K valley (dashed curve
with crosses) but decreases in the K$'$ one (dashed curve with dots).
When the intervalley hole-phonon scattering is included, 
even at low temperature, the difference of the SRTs is markedly suppressed,
thanks to the enhanced intervalley scattering by emitting phonons as mentioned
above.

\subsubsection{Valley polarization}

\begin{figure}[htb]
  {\includegraphics[width=8cm]{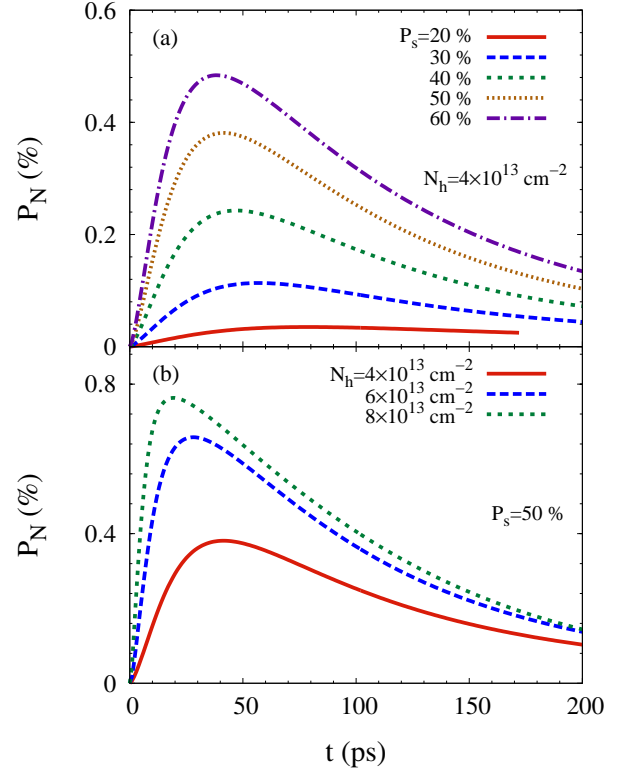}}
  \caption{(Color online) The temporal evolution of the valley polarization
    $P_N$ with different hole densities $N_h$ and initial spin polarizations
    $P_s$ at $T=50~$K.}  

\label{figyw10}
\end{figure}
At large spin polarization and low temperature,  it has been understood that the
faster spin relaxation in the K valley makes the density for spin-down (-up)
holes larger (smaller) than that
with the same spin in the K$'$ valley, triggering the intervalley scattering of
spin-down (-up) holes from the K (K$'$) to the K$'$ (K) valley by emitting
phonons, as shown in Fig.~\ref{figyw1}. 
During this process, it is discovered that the initial equal
occupations of holes in the two valleys are broken, leading to the build up of valley
polarization. To realize the large difference of the
spin relaxation processes with $\tau^{\rm
  K}_s{\ll}\tau^{\rm K'}_s$,  the electric field in our calculation satisfies
$\eta{E_{z}}=-{\Omega_{\rm HF}}|_{t=0}$ for given hole density and
initial spin polarization, and then the temporal evolution of the valley 
polarization $P_N=(N_{\rm K'}-N_{\rm K})/{N_{h}}$ at different hole densities
and initial spin polarizations when $T=50~$K 
are plotted in Fig.~\ref{figyw10}. It is seen that over
time, the valley polarization first increases and then decreases after reaching
the maximum. This temporal dependence can be understood as follows. 

\begin{figure}[htb]
  {\includegraphics[width=9cm]{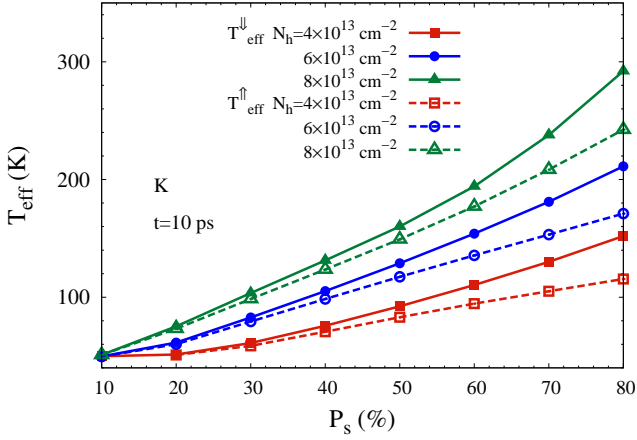}}
\caption{(Color online) The effective hot-hole temperatures versus the initial
  spin polarization $P_s$ at different hole densities when $t=10~$ps. Solid
  (Dashed) curves: for spin-down (-up) holes in the K valley. }   
\label{figyw11}
\end{figure}

By using the hot-hole Fermi distribution since the hole distribution in the
temporal evolution exhibits as a quasi hot-hole Fermi distribution [see the inset in 
Fig.~\ref{figyw8}(b)], the temporal evolution of $P_N$ in the
degenerate limit can be written as [given 
in Eq.~(\ref{vpfk5}) in Appendix~\ref{DD}]   
\begin{eqnarray}
\label{P_Nrate}
\frac{{\partial}P_N}{{\partial}t}=\frac{D_s}{N_h}(\frac{\Delta{E^{\Downarrow}}}{\tau^{{\rm
      K}\rightarrow{\rm K'}}_{p~\Downarrow}}-\frac{\Delta{E^{\Uparrow}}}{\tau^{{\rm 
    K'}\rightarrow{\rm K}}_{p~\Uparrow}}).
\end{eqnarray} 
Here, with ${\Delta}E^{\Downarrow
  (\Uparrow)}=\Omega_{\xi}-{\Delta}E_F^{\Downarrow (\Uparrow)}$, the intervalley
scattering time $\tau^{{\rm 
    K(K')}\rightarrow{\rm K'(K)}}_{p~\Downarrow (\Uparrow)}$ of spin-down (-up)
holes is given in Eq.~(\ref{intertime}), but substituting $T$ with the effective hot-hole
temperature $T^{\Downarrow (\Uparrow)}_{\rm eff}$ of spin-down (-up)
holes. 
   
At the beginning of the temporal evolution,  
$\Delta{E_F^{\Downarrow (\Uparrow)}}|_{t=0}=0$, and
hence the intervalley scattering rate by emitting phonons $1/\tau^{{\rm
    K(K')}\rightarrow{\rm K'(K)}}_{p~\Downarrow (\Uparrow)}{\approx}0$
[Eq.~(\ref{intertime})] whereas the one by absorbing phonons is 
negligible at low temperature. Therefore, the 
out-of-plane spins relax independently in the two valleys with different SRTs
$\tau^{\rm K}_{sz}$ and $\tau^{\rm K'}_{sz}$ [Eq.~(\ref{tau_s})].
Over time, due to the difference between $\tau^{\rm
  K}_{sz}$ and $\tau^{\rm K'}_{sz}$, $\Delta{E_F^{\Downarrow (\Uparrow)}}$
[Eq.~(\ref{difference})] increases with $\Delta{E_F^{\Uparrow}}=\Delta{E_F^{\Downarrow}}$, and then $1/\tau^{{\rm
    K(K')}\rightarrow{\rm   
    K'(K)}}_{p~\Downarrow (\Uparrow)}$ becomes larger, meaning that the intervalley
scattering by emitting phonons is triggered. Moreover, it is found that
    $\tau^{{\rm K}\rightarrow{\rm K'}}_{p~\Downarrow}<\tau^{{\rm K'}\rightarrow{\rm K}}_{p~\Uparrow}$, which
arises from the larger effective hot-hole temperature for spin-down holes than
that for spin-up ones. Specifically,
 the effective hot-hole temperatures 
for spin-down and -up holes in the K valley at different hole densities and initial spin
polarizations when $t=10~$ps are plotted in Fig.~\ref{figyw11}.  $T^{\Uparrow
  (\Downarrow)}_{\rm eff}$ is found to increase with increasing the hole density or spin
polarization due to the larger Fermi energy difference between spin-up and -down
holes and the enhanced electric field ($|\eta{E_{z}}|=|{\Omega_{\rm
    HF}}|_{t=0}$). Furthermore, it is seen that $T^{\Downarrow}_{\rm eff}$ is
always larger than $T^{\Uparrow}_{\rm eff}$, which arises from that the spin
precession at  
large spin polarization brings more spin-down holes occupying the states with 
the energies higher than the corresponding Fermi energy. Similar behaviors can
also be observed in the K$'$ valley. Consequently, with
 $\tau^{{\rm K}\rightarrow{\rm
    K'}}_{p~\Downarrow}<\tau^{{\rm K'}\rightarrow{\rm K}}_{p~\Uparrow}$ and
$\Delta{E_F^{\Uparrow}}=\Delta{E_F^{\Downarrow}}$, one has 
${\partial}_tP_N>0$ in Eq.~(\ref{P_Nrate}), leading to the increase of the
valley polarization at the first tens of picosecond. 

With further increase of time, the spin
polarization (i.e., the HF effective magnetic field) becomes smaller due to the spin relaxation, 
leading to the decrease of the difference between $\tau^{\rm K}_{sz}$ and
$\tau^{\rm K'}_{sz}$ by reducing the difference between $|\Omega^{\rm K}_{\rm eff}|$ and
$|\Omega^{\rm K'}_{\rm eff}|$. Then ${\Delta}E_F^{\Uparrow
  (\Downarrow)}$ induced by different $\tau^{\rm
  K}_{sz}$ and $\tau^{\rm K'}_{sz}$ at the first tens of picosecond is
suppressed by the intervalley scattering,  and hence $P_N$ decreases since
$P_NN_h=D_s(\Delta{E_F^{\Uparrow}}-\Delta{E_F^{\Downarrow}})$.  Moreover,
with $P_N>0$, $\Delta{E_F^{\Uparrow}}$ is relatively
larger than $\Delta{E_F^{\Downarrow}}$, and hence one has $\tau^{{\rm K'}\rightarrow{\rm
        K}}_{p~\Uparrow}<\tau^{{\rm
        K}\rightarrow{\rm K'}}_{p~\Downarrow}$ from Eq.~(\ref{intertime}) since the 
    difference between
    $T^{\Uparrow}_{\rm eff}$ and $T^{\Downarrow}_{\rm eff}$ (see
    Fig.~\ref{figyw11}) becomes smaller at the smaller spin polarization over
    time. Consequently, with the fact that the difference 
    between $\tau^{{\rm K}\rightarrow{\rm
        K'}}_{p~\Downarrow}$ and $\tau^{{\rm K'}\rightarrow{\rm K}}_{p~\Uparrow}$
    makes the dominant contribution in Eq.~(\ref{P_Nrate}),  one has
    ${\partial}_tP_N<0$, leading to the decrease of the valley polarization after reaching the maximum.

\begin{figure}[htb]
  {\includegraphics[width=8.8cm]{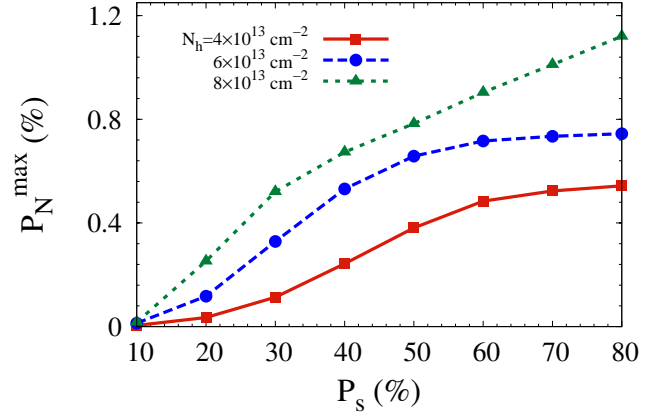}}
\caption{(Color online) The maxima of the valley polarization $P_N^m$ versus the initial
  spin polarization $P_s$ at different hole densities. }   
\label{figyw12}
\end{figure}

The maxima of the valley polarization $P^m_N$ are plotted in
Fig.~\ref{figyw12} at different hole densities and initial spin
polarizations. It is discovered that this maximum increases with
increasing the hole density and/or initial spin polarization. 
This dependence can be understood from the following analysis. As
demonstrated above, the intervalley hole-phonon 
scattering suppresses the difference of the SRTs in the two valleys [see
Fig.~\ref{figyw7}(c)]. By taking into account this suppression, from the KSBEs,
the maximum $P^m_N$ of the valley polarization is written as 
[given in Appendix~\ref{DD}]     
\begin{eqnarray}
\label{fvp}
P^m_N=\left[\frac{F}{2}\frac{\tau_p^*}{N_h}\left(\frac{N^{\Uparrow}_{\rm
      K}-N^{\Downarrow}_{\rm K}}{\tau^{\rm K}_{sz}}-\frac{N^{\Uparrow}_{\rm
      K'}-N^{\Downarrow}_{\rm K'}}{\tau^{\rm K'}_{sz}}\right)\right]\big|_{t=t_m}.
\end{eqnarray}
Here, the prefactor $F={(e^{\beta^{\Uparrow}_{\rm eff}\Omega_{\xi}}/\beta^{\Uparrow}_{\rm eff}-e^{\beta^{\Downarrow}_{\rm eff}\Omega_{\xi}}/\beta^{\Downarrow}_{\rm eff})}/\Omega_{\xi}$; $t_m$ is the time when
the valley polarization reaches the maximum.  To elucidate the
trend of $P^m_N$ with increasing initial spin polarization and hole density,
we consider $t_m\approx0$ approximately since
$t_m$ (around tens of picosecond) 
is much smaller than the SRT (larger than hundreds of picosecond). Then with
${\Omega^{\rm K'}_{\rm eff}}|_{t=0}=2{\eta}E_{z}$ and 
${\Omega^{\rm K}_{\rm eff}}|_{t=0}=0$ in Eq.~(\ref{tau_s}), an estimation of
$P^m_N$ can be written as 
\begin{eqnarray}
\label{pm}
P^m_N&\approx&{F|_{t=t_m}}\frac{P_s}{4}\frac{\tau_p^*}{\tau^{\rm K}_{sz}}\left(1-\frac{1}{1+4|\eta{E_z}\tau_p|^2}\right). 
\end{eqnarray}
Therefore, with the increase of hole density or
initial spin polarization and hence the electric field
($\eta{E_{z}}=-{\Omega_{\rm HF}}|_{t=0}$), $\tau^{\rm K}_{sz}$ becomes faster,
and $P^{m}_N$ tends to increase.

\section{SUMMARY}
\label{summary}

In summary, we have investigated the hole spin relaxation due to the Rashba SOC 
induced by an external perpendicular electric field in BL WSe$_{\bm 2}$, with
all the relevant scatterings included. The Rashba SOC in BL WSe$_{\bm 2}$ is
constructed from the ML Rashba SOC in each layer. Different from
ML Rashba SOC, the out-of-plane component of the BL Rashba SOC 
acts as opposite Zeeman-like fields in the 
two valleys.  For in-plane spins, this Zeeman-like field, opens an intervalley
spin relaxation channel in the presence of the intervalley hole-phonon
scattering, similar to the case of ML
MoS$_2$\cite{s_4,s_5} due to the intrinsic SOC. 
For 
out-of-plane spins, this Zeeman-like field is superimposed by the identical HF
effective magnetic fields in the two valleys, and hence the valley with same
(opposite) directions between these two fields has a larger (smaller) total
effective magnetic field. Situations of weak and strong HF effective magnetic
fields (i.e., small and large spin polarizations) are both considered. This
difference of the total effective magnetic fields between two valleys greatly
enriches the physics of the spin and valley dynamics in BL WSe$_2$.

In the case of small out-of-plane spin polarization, the total
effective magnetic field $\Omega^{\mu}_{\rm eff}$ is determined by the
Zeeman-like field ${\mu}{\eta}E_z$. We find
that the intervalley 
hole-phonon scattering is marginal to the  spin relaxation,
indicating that the 
out-of-plane spins relax independently in the two valleys. Identical SRTs in
the two valleys are obtained. We find that the behavior of the SRT
is very sensitive to the electric field,  and the different $\tau_p$
    dependences of the SRT are observed at different electric fields.
Specifically, the SRT can be divided into two regimes: regime
I, the anomalous EY-like regime ($\tau_{sz}\propto\tau_p$) when $|\Omega_{\rm
  eff}\tau_p|>1$;
regime II, the normal strong scattering regime ($\tau_{sz}\propto\tau^{-1}_p$)
when $|\Omega_{\rm eff}\tau_p|<1$.   
In regime II, the peak in the temperature
(density) dependence due to the crossover from the degenerate to the
nondegenerate limit is observed, similar to the studies in
semiconductors.\cite{pw_a,pw_2,pw_4,pw_5,pw_6,pw_8}    
By increasing the electric field (i.e., total effective magnetic field) to a
medium one ($|\Omega_{\rm eff}\tau_p|{\approx}1$), the crossover between regimes I and II is
observed. At this medium field,
the peak in   
the temperature dependence vanishes when the crossover between regimes I and II
is close to the crossover from the degenerate to the nondegenerate limit,
whereas the peak in the hole density dependence is always observed. 
This arises from the different $\tau_p$ dependences of the SRT between
regimes I and II. 

In the case of large out-of-plane spin polarization (i.e., strong HF effective
magnetic field), a large difference of the total effective magnetic fields
between two valleys is obtained. Due to this difference,  different
out-of-plane SRTs in the two valleys are observed when the intervalley
hole-phonon scattering is weak at low temperature and low hole
density. Specifically, the SRT in the valley with the smaller total effective
magnetic field is faster than that in the other valley. Moreover, by tuning the
Zeeman-like and HF effective magnetic fields, the valley with the larger total
effective magnetic field can fall into regime I ($|\Omega_{\rm eff}\tau_p|>1$)
while the other valley in regime II ($|\Omega_{\rm eff}\tau_p|<1$),
and hence the different behaviors of the SRTs between regimes I and II can be
observed in the different valleys simultaneously. This difference of
the SRTs between two valleys can be markedly suppressed through enhancing the
intervalley hole-phonon scattering by increasing temperature or hole density.
Additionally,  at low temperature in the degenerate limit, due to the
weak intravalley hole-phonon scattering but relatively strong hole-hole Coulomb
scattering, the fast spin precessions result in
a quasi hot-hole Fermi distribution characterized by an effective hot-hole
temperature $T_{\rm eff}$ larger than $T$, which also enhances the intervalley
hole-phonon scattering to suppress the difference of the SRTs.  

It is interesting to discover that at large spin polarization and low
temperature in the degenerate limit,  the initially equal hole
densities in the two valleys are broken in the temporal evolution, and a
valley polarization is built up. This arises from the different spin
relaxation processes in the two valleys at large spin polarizations.
Specifically, in the temporal evolution, the different spin relaxation processes
lead to different hole densities of each spin between the K and the K$'$ valleys,
but with opposite signs between spin-up and -down holes. Therefore,  the
spin-conserving intervalley hole-phonon scattering,  which transfers holes from
the high density valley into the low density one, has opposite scattering
directions between spin-down and -up holes.  In addition, the effective hot-hole
temperatures for spin-up and -down holes are found to be different,  leading to
different spin-conserving intervalley scattering rates. Consequently, with the
different intervalley scattering rates in the two opposite scattering
directions,  the initial equal occupations of holes in the two valleys are
broken, inducing the valley polarization.

Different from the out-of-plane spin relaxation where the
intervalley hole-phonon scattering is marginal at small spin polarization but makes an
important contribution at large one, for in-plane spin relaxation,
the SRT is found to be insensitive to the spin polarization and
the intervalley hole-phonon scattering always makes the dominant
contribution  even
at low temperature.  This arises from the
intervalley spin relaxation channel induced by the Zeeman-like field in the
presence of the
intervalley hole-phonon scattering, similar to the case of ML
MoS$_2$,\cite{s_4} and this intervalley channel 
dominates the in-plane hole spin relaxation in BL WSe$_2$.

\begin{acknowledgments}
This work was supported by the National Natural Science Foundation of
China under Grant No.\ 11334014, the National Basic Research Program of China
under Grant No.\ 2012CB922002, and the Strategic Priority Research Program of
the Chinese Academy of Sciences under Grant No.\ XDB01000000. One of the authors
(M.W.W) would like to thank Takashi Kimura at Kyushu University for hospitality
where the work is finished. We acknowledge Y. Song and H. Dery for providing the
identification of the $K_6$ phonon modes in the phonon energy spectrum.

\end{acknowledgments}

\begin{appendix}
\section{Rashba SOC in BL TMDs}
\label{AA}
Due to the weak interlayer coupling in BL TMDs, we use the ML Rashba SOC\cite{R}
in each layer to construct the BL Rashba SOC.  
The ML Rashba SOC by Korm{\'a}nyos {\em et al}.\cite{R} can be written
as \begin{equation} 
H_{\rm Rashba}^{\rm ML}=\gamma_{R}(k_x\sigma_y-k_y\sigma_x)E_z, 
\end{equation}
where $\sigma_{j}$ ($j=x,y$) are the Pauli matrices for the real spin; the ML
Rashba coefficients
$\gamma_{R}$ can be extracted from the electric field dependence of the
effective mass in the work by Zibouche {\em et al}.,\cite{A_2} which are given
in Table~\ref{Rp}. 

\begin{table}[htb]
  \caption{Rashba parameters of electrons $\gamma_R^e$ and holes $\gamma_R^h$ for ML MoS$_2$,
WS$_2$, MoSe$_2$ and WSe$_2$, with the unit being {\r A}$^2$.} 
\label{Rp}
  \begin{tabular}{l l l l l}
    \hline
    \hline
    &\;\;\;\;\;MoS$_2$\;\;\;&\;\;\;\;WS$_2$\;\;\;&\;\;\;\;MoSe$_2$\;\;\;&\;\;\;WSe$_2$\\  
    \hline
    $\gamma_R^e$&\;\;\;\;$0.0030$&\;\;\;$0.0326$&\;\;\;\;$0.0110$&\;\;\;$0.0416$\\
    $\gamma_R^h$&\;\;\;\;$0.0410$&\;\;\;$0.1072$&\;\;\;\;$0.0422$&\;\;\;$0.1166$\\
    \hline
    \hline
\end{tabular}
\end{table}

In BL TMDs, the Hamiltonian of the lowest four hole bands
    in the two layers near the K
(K$'$) point by Gong {\em et
  al}.\cite{Hamiltonian1} reads
\begin{equation}
H_{v}^{\rm BL}=\varepsilon_{\mu{\bf
    k}}+\lambda_v{(1+{\mu}{\tau}_z\sigma_z)}/{2}+t_{{\perp}k}\tau_x.
\end{equation}
Here, $\lambda_v$ is the spin splitting of the lowest two hole bands in each
layer; $t_{{\perp}k}=t_{\perp}(1-{a^2t^2k^2}/{\Delta^2})$ with $t_{\perp}$ and $t$
representing the interlayer and nearest-neighbor intralayer hopping,
respectively;  $\tau_{i}$ stands for the Pauli matrices for layer pseudospin; 
$a$ and $\Delta$ are the lattice constant and band gap, respectively. The
specific values of these parameters 
are taken from
Ref.~\onlinecite{Hamiltonian1}.

In the presence of an external perpendicular electric field, the total hole Hamiltonian near the K
(K$'$) point in BL TMDs reads
\begin{eqnarray}
H_{v}&=&\varepsilon_{\mu{\bf
    k}}+\lambda_v{(1+{\mu}\tau_z\sigma_z)}/{2}+t_{{\perp}k}\tau_x \nonumber \\
&&+\big[{\eta}\tau_z+\gamma_{R}(k_x\sigma_y-k_y\sigma_x)\big]E_z,    
\end{eqnarray}
where the electric dipole coefficient $\eta$
can be obtained from the energy splitting of the lowest hole bands in the work by
Zibouche {\em et al}..\cite{A_1}    

By treating the ML Rashba SOC and the interlayer hopping as
perturbations due to the large energy splitting $\lambda_v$ in each layer, we
construct the effective Hamiltonian by keeping the lowest two hole bands in two
layers through the L{\"o}wdin partition method.\cite{dia_1,dia_2} Up to the
third order of the momentum, the effective hole Hamiltonian, which is consistent
with the work by Yuan {\em et al}.\cite{electric2} is written as
\begin{equation}
H_{v}=\varepsilon_{\mu{\bf
    k}}+E_{z}[\nu_{k}(k_x\sigma_y-k_y\sigma_x)+\mu{\eta}\sigma_{z}],  
\end{equation}
where $\nu_{k}=\nu(1+{\alpha}k^2)$ with $\nu={\gamma_{\rm
  R}t_{\perp}}/{\lambda_v}$ and $\alpha=-{a^2t^2}/{\Delta^2}$.

\section{Derivation of Eq.~(\ref{tau_s})}
\label{CC}
We derive Eq.~(\ref{tau_s}) based on the KSBEs in the presence of the out-of-plane
total effective magnetic field (along the ${\hat z}$-axis) with only the
long-range hole-impurity scattering included. In the derivation,
we transform the KSBEs into the
interaction picture, and further use the strong-scattering approximation. 
  
The KSBEs in the collinear space with only the
long-range hole-impurity scattering included can be written as  
\begin{eqnarray}
\label{ts1}
&\partial_{t}\rho_{\mu{\bf k}}&=i\nu_{k}\big[k_y\sigma_x/2-k_x\sigma_y/2,\rho_{\mu{\bf
  k}}\big]-i\Omega^{\mu}_{\rm eff}\big[\sigma_z/2,\rho_{\mu{\bf k}}\big]\nonumber\\
&&-{2{\pi}N_i}\sum_{\bf
  k'}|V_{{\bf k}-{\bf k'}}|^2\delta(\varepsilon_{\bf k}-\varepsilon_{\bf k'})(\rho_{\mu{\bf k}}-\rho_{\mu{\bf k'}}).
\end{eqnarray}
By transforming the density matrix into the interaction picture as
\begin{eqnarray}
 \widetilde{\rho}_{\mu{\bf
  k}}&=&\exp({i\Omega^{\mu}_{\rm eff}\sigma_z/2})\rho_{\mu{\bf k}}\exp({-i\Omega^{\mu}_{\rm
    eff}\sigma_z/2}),
\end{eqnarray}
the KSBEs in the interaction picture become 
\begin{eqnarray}
&\partial_{t}\widetilde{\rho}_{\mu{\bf
  k}}=&i\nu_{k}\big[k_y\widetilde{\sigma}_x/2-k_x\widetilde{\sigma}_y/2,\widetilde{\rho}_{\mu{\bf
  k}}\big]-{2{\pi}N_i}~~~~~~\nonumber\\ 
&&\times\sum_{\bf
  k'}|V_{{\bf k}-{\bf k'}}|^2\delta(\varepsilon_{\bf k}-\varepsilon_{\bf k'})(\widetilde{\rho}_{\mu{\bf
  k}}-\widetilde{\rho}_{\mu{\bf k'}}), 
\end{eqnarray}
with $\widetilde{\sigma}_{i}=\exp({i\Omega^{\mu}_{\rm eff}\sigma_z/2})\sigma_{i}\exp({-i\Omega^{\mu}_{\rm
    eff}\sigma_z/2})$.
After the Fourier transformation 
\begin{eqnarray}
 \widetilde{\rho}^{l}_{\mu{\bf
  k}}=\frac{1}{2\pi}{\int}^{2\pi}_{0}d\theta_{\bf k}\widetilde{\rho}_{\mu{\bf
  k}}\exp(-il\theta_{\bf k}),
\end{eqnarray}
one gets
\begin{eqnarray}
{\partial_t\widetilde{\rho}_{\mu{\bf k}}^{l}}=-{\nu_kk}/4\big(\big[\widetilde\sigma_+,\widetilde{\rho}^{l-1}_{\mu{\bf k}}\big]-\big[\widetilde\sigma_-,\widetilde{\rho}^{l+1}_{\mu{\bf k}}\big]\big)-{\widetilde{\rho}^{l}_{\mu{\bf k}}}/{\tau_{k,l}},~~~
\end{eqnarray}
with
\begin{eqnarray}
 \frac{1}{\tau_{k,l}}=\frac{m^*N_i}{2\pi}\int^{2\pi}_{0}d\theta_{\bf q}|V_{{\bf q}}|^2(1-{\cos}l\theta).
\end{eqnarray}
Further keeping terms $|l|{\le}1$ and defining the spin vector, 
\begin{eqnarray}
\widetilde{\bf S}^l_{\mu{\bf k}}={\rm Tr}\big[\widetilde{\rho}_{\mu{\bf k}}^{l}{\bm {\sigma}}\big],
\end{eqnarray}
one obtains 
\begin{eqnarray}
\label{seq}
&&\big[\tau_{k,1}^2\partial_{t}^3+2\tau_k^1\partial_{t}^2+(1+|\Omega^{\mu}_{\rm
  eff}\tau_{k,1}|^2)\partial_{t}+\nu_k^2k^2\tau_{k,1}^2\partial_{t}\nonumber\\
&&~+{\nu_k^2k^2\tau_{k,1}}\big]\widetilde{S}^0_{\mu{\bf k}z}(t)=i\Omega^{\mu}_{\rm eff}(\nu_k^2k^2\tau_{k,1}^2).
\end{eqnarray}
In the strong scattering limit ($|\nu_kk\tau_{k,1}|^2{\ll}1$), Eq.~(\ref{seq})
becomes 
\begin{eqnarray}
\big[2\tau_{k,1}\partial_{t}^2+(1+|\Omega_{\rm
  eff}^{\mu}\tau_{k,1}|^2)\partial_{t}+\nu_k^2k^2\tau_{k,1}\big]\widetilde{S}^0_{\mu{\bf
    k}z}(t)=0.~~~~~
\end{eqnarray}
With the initial condition $\partial_{t}\widetilde{S}^0_{\mu{\bf k}z}(0)=0$, the
solution of $\widetilde{S}^0_{\mu{\bf k}z}(t)$ is given by 
\begin{widetext}
\begin{eqnarray}
\widetilde{S}^0_{{\bf k}z}(t)=\frac{\widetilde{S}^0_{{\bf k}z}(0)}{2}\left\{(1+\frac{1}{\sqrt{1-c_z^2}})\exp\left[-{\frac{t(1-\sqrt{1-c_z^2})}{4\tau_{k,1}/(1+|\Omega^{\mu}_{\rm
  eff}\tau_{k,1}|^2)}}\right]+(1-\frac{1}{\sqrt{1-c_z^2}})\exp\left[-{\frac{t(1+\sqrt{1-c_z^2})}{4\tau_{k,1}/(1+|\Omega^{\mu}_{\rm
  eff}\tau_{k,1}|^2)}}\right]\right\},~~~~
\end{eqnarray}
\end{widetext}
with $c_z=2\sqrt{2}\nu_kk\tau_{k,1}/(1+|\Omega^{\mu}_{\rm
  eff}\tau_{k,1}|^2)$. In the
strong scattering limit ($c_z{\ll}1$), one has
\begin{eqnarray}
\label{psps}
\widetilde{S}^0_{\mu{\bf k}z}(t)={\widetilde{S}^0_{\mu{\bf k}z}(0)}\exp\big(-\frac{t}{8\tau_{k,1}/c_z^2}\big).
\end{eqnarray}
The out-of-plane SRT is therefore
\begin{eqnarray}
\label{taueq}
\tau^{\mu}_{sz}\approx(1+|\Omega^{\mu}_{\rm eff}\tau_{k,1}|^2)/(\nu_k^2k^2\tau_{k,1}).
\end{eqnarray}
When all the relevant intravalley scatterings are included, by replacing
$\tau_{k,1}$ with $\tau_p$, Eq.~(\ref{taueq}) becomes Eq.~(\ref{tau_s}).

Moreover, without the intervalley scattering included, from Eq.~(\ref{psps}),
the density difference of
holes of each spin [$N_{\mu}={\rm Tr}(\rho_{\mu{\bf k}})$] between the two valleys induced by the different SRTs can be
written into
\begin{equation}
\label{differenceCC}
{\Delta}N_h^{\Uparrow(\Downarrow)}={0.25P_sN_h}(e^{-t/{\tau^{\rm
      K'}_{sz}}}-e^{-t/{\tau^{\rm K}_{sz}}}).
\end{equation}

\section{The  hole-phonon scattering matrix elements}
\label{BB}

For the intravalley hole-phonon scattering, a symmetry group analysis of
lattice vibrations at the $\Gamma$ point, which belongs to the space group ${\rm
  D}_{6h}$, has been performed in the previous work.\cite{ph_v1} 
The decomposition into irreducible representations is as follows: 
\begin{eqnarray}
\Gamma&=&A_{\rm 1g}{\oplus}2A_{\rm 2u}{\oplus}B_{\rm 1u}{\oplus}2B_{\rm
  2g}~\nonumber \\ 
&&{\oplus}E_{\rm 1g}{\oplus}2E_{\rm 1u}{\oplus}E_{\rm
  2u}{\oplus}2E_{\rm 2g}.
\end{eqnarray}
The vibration pattern of these phonon modes can be found in
Refs.~\onlinecite{ph_v1,ph_v2,ph_v3,ph_v4}.
It is noted that only the modes [including in-plane OP ($E_{\rm 1u},E^{1,2}_{\rm
  2g}$) and AC ($E^2_{\rm 1u}$) phonons and out-of-plane 
OP ($A^2_{\rm 2u}$, $B^{1,2}_{\rm 2g}$) and AC ($A^1_{\rm 2u}$) phonons], which induce the
vibrations of the transition metal atoms, can trigger the hole-phonon 
scattering. 

Due to the weak interlayer coupling in BL TMDs,\cite{ph_v1} the BL phonon
vibrations are thought to consist of the corresponding ML phonon vibrations in
each layer, and the BL phonon energy
spectrum is close to the corresponding ML one.\cite{Thick} Therefore, for the
in-plane phonons,  
the BL intravalley hole-phonon scattering matrix elements are constructed by using
the ML ones, which have been reported in the work by Jin {\em et al.}.\cite{ph_6}
By considering the number of the in-plane phonon modes in BL TMDs, the
scattering matrix elements are given by     
\begin{eqnarray} 
&&|{M^{{\rm AC},E^{1}_{\rm 2g}}_{\mu\mu'{\bf
      q}}}|^2=\frac{(\Xi)^2q}{2{\rho}v_{\rm
    LA}}\delta_{\mu',\mu},\\    
&&|{M^{E_{\rm 1u},E^{1}_{\rm 2g}}_{\mu\mu'{\bf
      q}}}|^2=\frac{(D_{\rm OP})^2}{{\rho}\Omega_{\Gamma,{
      E_{\rm 1u}}}}\delta_{\mu',\mu}.
\end{eqnarray}
But for the out-of-plane phonons, it has been shown that the contribution of the
out-of-plane phonons in the hole-phonon scattering is marginal in ML
TMDs.\cite{ph_6} However, the out-of-plane OP phonons ($A^2_{\rm
  2u}, B^1_{\rm 2g}$) in BL TMDs, which induce the relative out-of-plane vibrations of the
transition metal atoms in the two layers, can largely influence the variety of
the interlayer hopping, triggering the intravalley hole-phonon scattering.
By using the tight-binding model according to the arXiv version of the work by
Viljas and Heikkil{\"a},\cite{ph_a} we derived the matrix elements of this kind
of hole-phonon scattering 
\begin{eqnarray} 
&&|{M^{B^{2}_{\rm 2g}}_{\mu\mu'{\bf
      q}}}|^2=\frac{{t^{\prime}_{\perp}}^2}{{\rho}{\gamma}q^2}\delta_{\mu',\mu},\\      
&&|{M^{A^2_{\rm 2u},B^{1}_{\rm 2g}}_{\mu\mu'{\bf
      q}}}|^2=\frac{{t^{\prime}_{\perp}}^2}{2{\rho}\Omega_{\Gamma,{ 
      A^2_{\rm 2u}}}}\frac{2M_d}{M_t}\delta_{\mu',\mu}.
\end{eqnarray}

For the spin-conserving intervalley hole-phonon scattering in BL TMDs, with the
intervalley hole-phonon scattering in each layer suppressed due to the large energy
splitting,\cite{s_1} holes in a given valley at a given layer are scattered into
the other valley at different layer. Only the intervalley phonons, which induce
the relative out-of-plane vibrations of the transition metal atoms in the two
layers and hence lead to the variety of the interlayer hopping, can trigger this
kind of the intervalley hole-phonon scattering. To derive the intervalley
hole-phonon scattering matrix
elements, one needs to know the intervalley phonon vibrations. As
mentioned above, the BL phonon vibrations can be constructed by
using the ML ones.      
Through the group theory analysis of ML lattice vibrations at the K point, 
which belongs to the space group ${\rm C}_{3h}$, the decomposition into
irreducible representations is as follows:   
\begin{eqnarray}
{\rm K}=2A'{\oplus}2E_1'{\oplus}E_2'{\oplus}A{''}{\oplus}E_1''{\oplus}2E_2''.  
\end{eqnarray}
Only the $K_6$ phonon modes\cite{s_1} (correspond to the representation
$E^{\prime\prime}_{2}$) are the out-of-plane vibrational modes, and the vibrations
of the two branches of $K_6$ phonons ($K_6^{\rm H}$ and $K^{\rm L}_6$) should be
two kinds of orthogonal combinations of the out-of-plane vibration $\psi_t$ of the
transition metal atoms and the in-plane vibration $\psi_d$ of the dichalcogenide
atoms: 
\begin{eqnarray}
\psi_{K_6^{\rm L}}&=&\frac{\psi_d+A\psi_t}{\sqrt{M_{\rm W}+2A^2M_{\rm Se}}}, \\
\psi_{K_6^{\rm H}}&=&\frac{A\sqrt{2M_{\rm Se}/M_{\rm W}}\psi_d-\sqrt{M_{\rm
    W}/2M_{\rm Se}}\psi_t}{\sqrt{M_{\rm W}+2A^2M_{\rm Se}}}.
\end{eqnarray}
To obtain the specific combination $A$, one needs to solve the kinetic equation, which is
beyond the scope of this investigation.  We take $A=1$ approximately here.\cite{group}
Then, we construct BL intervalley phonon
vibrations by using the ML ones in each layer but with opposite vibration
directions between the two layers, and by using the tight-binding model
according to the same work by Viljas and Heikkil{\"a},\cite{ph_a} the
scattering matrix elements are given by  
\begin{eqnarray} 
&&|{M^{K_6^{\rm L}}_{\mu\mu'{\bf
      q}}}|^2=\frac{t^{\prime}_{\perp}}{2\rho\Omega_{K,K^{\rm L}_{6}}}\delta_{\mu',-\mu},\\ 
&&|{M^{K_6^{\rm H}}_{\mu\mu'{\bf
      q}}}|^2=\frac{t^{\prime}_{\perp}}{2\rho\Omega_{K,K^{\rm
      H}_{6}}}\frac{2M_{d}}{M_t}\delta_{\mu',-\mu}. 
\end{eqnarray}

\section{Valley polarization from the KSBEs}
\label{DD}
With the hole density in each valley $N_{\mu}={\rm Tr}(\rho_{\mu{\bf k}})$, by
taking trace from both sides of the KSBEs, one obtains  
\begin{eqnarray}
\label{vpfk1}
\partial_t{N_{\mu}}&=&{{\rm Tr}(\partial_{t}\rho_{\mu {\bf k}}|_{\rm
  coh})}+{\rm Tr}(\partial_{t}\rho_{\mu {\bf k}}|^{\rm intra}_{\rm scat})+{\rm Tr}(\partial_{t}\rho_{\mu {\bf k}}|^{\rm inter}_{\rm scat}).\nonumber\\
\end{eqnarray}
Since the hole density in each valley is not affected by the spin precessions
and intravalley scattering, one has ${{\rm Tr}(\partial_{t}\rho_{\mu {\bf
      k}}|_{\rm coh})}=0$ and ${\rm Tr}(\partial_{t}\rho_{\mu {\bf k}}|^{\rm 
  intra}_{\rm scat})=0$. Additionally, in our calculation, 
the energy-splitting induced by the Zeeman-like term $\mu{\eta}E_z$ is much smaller than the Fermi
energy, and hence by neglecting this energy splitting in
Eq.~(\ref{interscatt}), one has
\begin{widetext}
\begin{eqnarray}
\label{vpfk2}
{\partial}_tN_{\mu}={2\pi}\sum_{{\bf
  kk'q\sigma}\mu'}|M_{\mu\mu'}^{\xi}|^2\big\{\big[f^{\sigma}_{{\mu'}{\bf
    k+q}}(1-f^{\sigma}_{\mu{\bf k}})(N_{\bf  
    q}+1)-f^{\sigma}_{\mu{\bf k}}(1-f^{\sigma}_{\mu'{\bf k+q}})N_{\bf q}\big]\delta(\varepsilon_{\bf
  k+q}-\varepsilon_{\bf k}-\Omega_{\xi})\nonumber\\
+\big[f^{\sigma}_{\mu'{\bf k+q}}(1-f^{\sigma}_{\mu{\bf k}})N_{\bf 
    q}-f^{\sigma}_{\mu{\bf k}}(1-f^{\sigma}_{\mu'{\bf k+q}})(N_{\bf q}+1)\big]\delta(\varepsilon_{\bf
  k}-\varepsilon_{\bf k+q}-\Omega_{\xi})\big\}.
\end{eqnarray}
\end{widetext}
As the hole distribution exhibits a quasi hot-hole
Fermi distribution behavior [see the inset in Fig.~\ref{figyw8}(b)] in the
temporal evolution, we use the hot-hole Fermi distribution characterized by
$T_{\rm eff}$ 
in Eq.~(\ref{vpfk2}). In addition, one has $N_{\bf q}{\approx}0$ at low
temperature ($k_BT{\ll}\Omega_{\xi}$), denoting the intervalley hole-phonon
scattering through absorbing phonons is negligible. Therefore, with the larger
density of spin-down (-up) holes in the K (K$'$) valley, 
Eq.~(\ref{vpfk2}) in the degenerate limit becomes 
\begin{eqnarray}
\label{vpfk3}
\frac{{\partial}N_{\mu}}{{\partial}t}=\frac{{\mu}D_s}{2}(\frac{\Delta{E^{\Uparrow}}}{\tau^{{\rm 
    K'}\rightarrow{\rm K}}_{p~\Uparrow}}-\frac{\Delta{E^{\Downarrow}}}{\tau^{{\rm 
      K}\rightarrow{\rm K'}}_{p~\Downarrow}}),
\end{eqnarray} 
where
\begin{eqnarray} 
\label{vpfk4}
\tau^{{\rm K' (K)}\rightarrow{\rm
    K (K')}}_{p~\Uparrow
  (\Downarrow)}&=&\tau^*_p(e^{\beta^{\Uparrow (\Downarrow)}_{\rm
    eff}{{\Delta}E^{\Uparrow (\Downarrow)}}}-1), \nonumber \\
\end{eqnarray}
for ${\Delta}E^{\Uparrow (\Downarrow)}>0$. The first (second) term on the right
handside of Eq.~(\ref{vpfk3}) comes from the
contribution of the spin-conserving intervalley scattering of spin-up (-down)
holes. 

Substituting $N_{\rm K'}-N_{\rm K}=P_NN_h$ into Eq.~(\ref{vpfk4}), the
temporal evolution of the valley polarization reads
\begin{eqnarray}
\label{vpfk5}
\frac{{\partial}P_N}{{\partial}t}=\frac{D_s}{N_h}(\frac{\Delta{E^{\Downarrow}}}{\tau^{{\rm 
      K}\rightarrow{\rm K'}}_{p~\Downarrow}}-\frac{\Delta{E^{\Uparrow}}}{\tau^{{\rm 
    K'}\rightarrow{\rm K}}_{p~\Uparrow}}).
\end{eqnarray}

Next we derive Eq.~(\ref{fvp}) based on the KSBEs. Specifically, it has been
shown that in the degenerate limit, the difference of the SRTs in the two
valleys is suppressed by the intervalley hole-phonon scattering [see
Fig.~\ref{figyw8}(b)]. With the
contribution of the intravalley scatterings to the spin relaxation given in
Eq.~(\ref{psps}), by neglecting the dephasing parts of the intervalley
hole-phonon scattering since the intervalley hole-phonon scattering is in the
weak-scattering limit, the temporal evolution of spin polarization in each 
valley becomes
\begin{eqnarray}
\label{vpfk6}
\frac{{\partial}P^{\mu}_s}{{\partial}t}=-\frac{P^{\mu}_s}{\tau^{\mu}_{sz}}+\mu\frac{D_s}{N_h}(\frac{\Delta{E^{\Downarrow}}}{\tau^{{\rm 
      K}\rightarrow{\rm K'}}_{p~\Downarrow}}+\frac{\Delta{E^{\Uparrow}}}{\tau^{{\rm 
    K'}\rightarrow{\rm K}}_{p~\Uparrow}}).
\end{eqnarray}
Further considering ${\partial}_tP^{\rm
  K}_s{\approx}{\partial}_tP^{\rm K'}_s$ in Eq.~(\ref{vpfk6}), one has
\begin{equation}
\label{vpfk7}
\frac{2D_s}{N_h}(\frac{\Delta{E^{\Downarrow}}}{\tau^{{\rm 
      K}\rightarrow{\rm K}}_{p~\Downarrow}}+\frac{\Delta{E^{\Uparrow}}}{\tau^{{\rm 
    K'}\rightarrow{\rm K}}_{p~\Uparrow}})\approx\frac{P^{\rm K}_s}{\tau^{\rm
  K}_{sz}}-\frac{P^{\rm K'}_s}{\tau^{\rm K'}_{sz}}. 
\end{equation}
Additionally, $\partial_tP_N=0$ in Eq.~(\ref{vpfk5}) when the 
valley polarization reaches the maximum at $t=t_m$, and then by using 
Eqs.~(\ref{vpfk4}) and~(\ref{vpfk7}), one obtains      
\begin{eqnarray}
\frac{{\Delta}E^{\Uparrow (\Downarrow)}}{e^{\beta^{\Uparrow (\Downarrow)}_{\rm
    eff}{{\Delta}E^{\Uparrow (\Downarrow)}}}-1}=\frac{\tau^*_pN_h}{4D_s}(\frac{P^{\rm K}_s}{\tau^{\rm
  K}_{sz}}-\frac{P^{\rm K'}_s}{\tau^{\rm K'}_{sz}}).
\end{eqnarray}
Taking 
$\beta_{\rm eff}^{\Downarrow(\Uparrow)}{\Delta}E_F^{\Downarrow
  (\Uparrow)}{\ll}1$ and
$P_NN_h=D_s({\Delta}E_F^{\Uparrow}-{\Delta}E_F^{\Downarrow})$, the maximum of
the valley polarization [Eq.~(\ref{fvp})] is obtained.

\end{appendix}

\end{document}